\documentclass[rmp,twocolumn
				]{revtex4-1}

\usepackage{graphicx}
\usepackage{url}
\usepackage{amsmath}
\usepackage{amssymb}
\usepackage{bm}
\usepackage{dsfont}
\usepackage{graphicx}
\usepackage{subfigure}
\usepackage{dcolumn}

\newcommand{\rd}{\mathrm{d}}

\newcommand{\Jb}{\mathcal{J}}
\newcommand{\no}{\hat{n}}
\renewcommand{\ao}{\hat{a}}
\renewcommand{\aa}{\hat{a}^\dag}
\newcommand{\Ho}{\hat{H}}
\newcommand{\Uo}{\hat{U}}
\newcommand{\Vo}{\hat{V}}

\newcommand{\Qo}{\hat{Q}}
\newcommand{\la}{\langle}
\newcommand{\ra}{\rangle}
\newcommand{\lla}{\langle\!\langle}
\newcommand{\rra}{\rangle\!\rangle}

\newcommand{\br}{{\bm r}}

\newcommand{\bxi}{{\bm \xi}}
\newcommand{\bF}{{\bm F}}

\newcommand{\ba}{{\bm a}}
\newcommand{\bk}{{\bm k}}
\newcommand{\bq}{{\bm q}}

\newcommand{\be}{\begin{equation}}
\newcommand{\ee}{\end{equation}}
\newcommand{\bes}{\begin{eqnarray}}
\newcommand{\ees}{\end{eqnarray}}

\begin{document}

\title{Atomic quantum gases in periodically driven optical lattices}

\date{July 13, 2016}

\author{Andr\'e Eckardt}
\email{eckardt@pks.mpg.de}
\affiliation{Max-Planck-Institut f\"ur Physik komplexer Systeme, 
			\mbox{N\"othnitzer Str.\ 38, 01187 Dresden, Germany}}

\begin{abstract}
Time periodic forcing in the form of coherent radiation is a standard tool for the 
coherent manipulation of small quantum systems like single atoms. In the last years, 
periodic driving has more and more also been considered as a means for the coherent 
control of many-body systems. In particular, experiments with ultracold quantum gases in 
optical lattices subjected to periodic driving in the lower kilohertz regime have 
attracted a lot of attention. Milestones include the observation of dynamic 
localization, the dynamic control of the quantum phase transition between a bosonic 
superfluid and a Mott insulator, as well as the dynamic creation of strong artificial 
magnetic fields and topological band structures. This article reviews these recent experiments 
and their theoretical description. Moreover, fundamental properties of periodically driven
many-body systems are discussed within the framework of Floquet theory, including heating, 
relaxation dynamics, anomalous topological edge states, and the response to slow parameter 
variations. 
\end{abstract}

\maketitle

\tableofcontents
\section{Introduction}

While time-periodic forcing in the form of coherent radiation is a standard tool for the 
coherent manipulation of small quantum systems like single atoms, traditionally it plays 
much less of a role in the context of many-body systems. However, recent experiments 
with ultracold atomic quantum gases in optical lattices demonstrate that periodic 
forcing can also be a powerful tool for the coherent manipulation of \emph{many-body} 
states and their dynamics. These experiments include the control of ballistic expansion 
of Bose-Einstein condensates via periodic driving \cite{LignierEtAl07}, coherent resonant
AC-induced tunneling \cite{SiasEtAl08, IvanovEtAl08, AlbertiEtAl09, HallerEtAl10}, the 
dynamic control of the quantum phase transition between a bosonic Mott insulator and a 
superfluid \cite{ZenesiniEtAl09}, the creation of kinetic frustration \cite{StruckEtAl11},
artificial magnetic fields \cite{AidelsburgerEtAl11, StruckEtAl12, StruckEtAl13,
AidelsburgerEtAl13, MiyakeEtAl13, AtalaEtAl14, KennedyEtAl15} and topological band 
structures \cite{JotzuEtAl14, AidelsburgerEtAl15}, coherent band coupling
\cite{GemelkeEtAl05, BakrEtAl11,ParkerEtAl13, HaEtAl15}, as well the coherent control of 
interaction blockade by means of resonant forcing \cite{MaEtAl11, ChenEtAl11, BakrEtAl11}. It 
is the fact that ultracold quantum gases are extremely clean, very well isolated from their 
environment, and highly controllable in a time-dependent fashion that allowed for these recent 
advances.

On a theoretical level, the idea of controlling lattice systems by means of strong 
periodic forcing (beyond the regime of linear response) dates back to the work of 
\onlinecite{DunlapKenkre86}. They investigated the spreading of a localized particle in 
a tight-binding chain under the influence of a sinusoidal force $F_\omega\cos(\omega t)$.
The forcing was found to slow down the linear spreading of the wave function by a factor 
of $\Jb_0(dF_\omega/\hbar\omega)$, with lattice constant $d$ and $\Jb_m$ denoting the 
Bessel function of the first kind of order $m$. The possibility to tune this factor to 
zero, and thus completely suppress the dispersion of the wave packet, was termed
\emph{dynamic localization}. The effect was much later observed with a Bose-Einstein 
condensate in a shaken optical lattice \cite{LignierEtAl07, EckardtEtAl09,
CreffieldEtAl10} and in arrays of optical wave guides, where one spatial direction plays 
the role of time \cite{LenzEtAl03, LonghiEtAl06, IyerEtAl07, DreisowEtAl08, SzameitEtAl09,
SzameitEtAl10}. It can be understood in terms of an effective modification of the band 
width (or the tunneling matrix element) by the same factor \cite{Holthaus92}. Whereas 
this modification is exact in the infinite translational invariant chain, for large 
frequencies it still holds approximately if the translational symmetry is
broken.\footnote{In this high-frequency limit the phenomenon is equivalent to the 
effective modification of the Land\'e factor of an off-resonantly driven atomic spin 
\cite{HarocheEtAl70} and the effect of \emph{coherent destruction of tunneling} 
\cite{GrossmannEtAl91} in a driven two-level system \cite{Shirley65, GrossmannHaenggi92, 
GomezLlorentePlata92,GrifoniHaenggi98} observed in an atom-beam experiment
\cite{KierigEtAl08}.} 
This effect was used in a series of proposals for the AC control of quantum mechanical 
localization by effectively squeezing the tunneling parameter relative to the strength 
of an isolated defect \cite{HoneHolthaus93}, on-site disorder \cite{HolthausEtAl95}, or a 
quasiperiodic perturbation \cite{DreseHolthaus97b}. The last reference is also the 
first proposal for the application of such a coherent control scheme to a system of 
ultracold atoms in a driven optical lattice. More recently, it was argued that for 
large driving frequencies the effective modification of the tunneling remains 
approximately valid also in the presence of interactions, so that it should be possible 
to control also the interaction-driven localization transition from a bosonic superfluid 
to a Mott-insulating state \cite{EckardtEtAl05b}, an effect later observed 
experimentally \cite{ZenesiniEtAl09}.

An important concept for the coherent control of time-periodically driven quantum 
systems, also called \emph{Floquet systems}, is the \emph{Floquet Hamiltonian} $\Ho^F_{t_0}$. 
It is defined to reproduce the time evolution generated by the Hamiltonian $\Ho(t)=\Ho(t+T)$ 
over one driving cycle $T$,\footnote{Let us make a note on terminology: The Floquet 
Hamiltonian, as it is defined here, is a special case of an effective Hamiltonian $\Ho_F$ 
introduced in Eq.~(\ref{eq:HFgauge}) below [see Eq.~(\ref{eq:HfloquetHeff}) and the paragraph 
containing it]. It is also called ``effective Hamiltonian'' by some authors. Moreover, the term
``Floquet Hamiltonian'' is sometimes used to denote the operator $\Qo(t)=\Ho(t)-i\hbar\rd_t$ 
[Eq.~\ref{eq:Q}] acting in the space of time-periodic states, which will be denoted
``quasienergy operator'' here.} 
\be\label{eq:UHeff}
\Uo(t_0+T,t_0)\equiv\exp\Big({-\frac{i}{\hbar}T\Ho^F_{t_0}}\Big).
\ee
Here $\Uo(t_2,t_1)$ denotes the time evolution operator from time $t_1$ to time $t_2$. 
Thus, when looking at the time evolution in a stroboscopic fashion in steps of the driving 
period $T$, the system behaves effectively as if it was described by the \emph{time-independent}
Hamiltonian $\Ho^F_{t_0}$. The effect of dynamic localization has to be understood in this 
sense. 

The simple equation~(\ref{eq:UHeff}) suggests a general strategy for the controlled 
manipulation of quantum systems. By tailoring the Hamiltonian $\Ho(t)$ of a system and 
its periodic time dependence the physics of a Floquet Hamiltonian $\Ho^F_{t_0}$ with 
desired properties can be realized. This concept of \emph{Floquet engineering} becomes 
of practical relevance, provided three conditions are fulfilled:
\begin{itemize}
\item[(i)] The system allows for the implementation of a suitable time-periodic driving 
scheme.
\item[(ii)] The system is well isolated from its environment such that dissipative 
processes happen on a time scale much longer than the driving period $T$.
\item[(iii)] The Floquet Hamiltonian can be computed theoretically, at least within a 
suitable approximation valid on the experimentally relevant time scale, and takes a simple 
form that allows for a clear interpretation. 
\end{itemize}
The first two requirements make ultracold atomic quantum gases, which are well isolated 
from their environment and provide a great freedom for time dependent parameter control, 
an optimal platform for Floquet engineering.  

Based on this strategy it is also possible to endow a system with qualitatively new 
properties. A prime example is the creation of artificial gauge fields (magnetic fields 
or spin orbit coupling), which among others \cite{DalibardEtAl11, GalitskiSpielman13,
GoldmanEtAl14} can be accomplished using Floquet engineering. For that purpose
charge-neutral atoms in an optical lattice are driven in such a way that they behave 
effectively as if they had a charge coupling to a magnetic field or to their spin. 
Such a proposal for the realization of an artificial magnetic field was first 
made by \onlinecite{SoerensenEtAl05},
based on a sequence of overlapping pulses during each cycle where external potentials 
and the amplitudes of the tunneling matrix elements in both directions are switched on 
in an alternating fashion. Later, simpler schemes, relying solely on the modulation of
on-site potentials, were realized experimentally. This includes the effective creation 
of a topologically non-trivial band structure by means of circular forcing
\cite{OkaAoki09}, known as \emph{Floquet topological insulator} (see also the related 
work by \onlinecite{KitagawaEtAl10, LindnerEtAl11}). Originally proposed for electrons in 
irradiated graphene, it was realized with fermionic atoms in a circularly shaken 
honeycomb-like lattice \cite{JotzuEtAl14} [as well as in an optical wave-guide 
experiment \cite{RechtsmanEtAl13}]. Lattice shaking was also employed to create kinetic 
frustration and staggered magnetic fields in a triangular optical lattice
\cite{EckardtEtAl10, StruckEtAl11, StruckEtAl12, StruckEtAl13}. Finally, the effective 
creation of magnetic fields can be achieved in a square lattice where tunneling against 
strong potential offsets is resonantly induced by driving the system with a moving 
secondary lattice \cite{Kolovsky11, BermudezEtAl11}, as has been demonstrated experimentally 
with bosonic atoms \cite{AidelsburgerEtAl11, AidelsburgerEtAl13, MiyakeEtAl13,
AtalaEtAl14, KennedyEtAl15, AidelsburgerEtAl15}. 

Despite these experimental results, which proof the great success of Floquet engineering in 
atomic quantum gases, it would be misleading to state that the stroboscopic time evolution of 
periodically driven quantum systems simply corresponds to that of some effective autonomous
(\emph{i.e.} non-driven) system. Even though we can define a Floquet Hamiltonian $\Ho^F_{t_0}$, 
its properties are generally quite different from those of the time-independent Hamiltonians 
used to describe autonomous many-body systems. These differences result from the absence of 
energy conservation in the driven systems, which is reflected in the fact that the Floquet 
Hamiltonian is not defined uniquely by relation (\ref{eq:UHeff}), since the logarithm is 
multivalued. Namely, its eigenvalues, the \emph{quasienergies}, are determined modulo the 
energy quantum $\hbar\omega$ only; the quasienergy spectrum can be represented on a circle. 
Thus, when switching on a time-periodic perturbation, eigenstates of the unperturbed
time-independent Hamiltonian with energies separated by some integer multiple of $\hbar\omega$ 
appear to be degenerate and can, therefore, hybridize.  As a consequence, the eigenstates of
$\Ho^F_{t_0}$ can be coherent superpositions of unperturbed states of rather different energy. 
In the above-mentioned experiments such resonant coupling plays two different roles. On the 
one hand it is sometimes exploited to induce coherent tunneling against static potential 
offsets of integer multiples of $\hbar\omega$ and plays a major role for engineering desired 
system properties. On the other hand, it also causes heating. For the purpose of Floquet 
engineering such heating has to be suppressed on the experimentally relevant time scale by a 
suitable choice of parameters. Further fundamental differences between the (stroboscopic) 
dynamics of periodically driven quantum systems and that of autonomous systems include, for 
example, the possible emergence of anomalous topological edge states or heating in response to 
slow parameter variations (see section \ref{sec:FloquetPicture}). 

This article reviews the status of Floquet engineering in systems of ultracold atomic quantum 
gases in periodically driven optical lattices. For this purpose, we will first briefly 
summarize a few general properties of time-periodically driven quantum systems in section
\ref{sec:GenProp}. In section \ref{sec:experiments} we will then describe recent experiments 
and explain them in terms of a common language and using simple intuitive approximations. This 
section addresses also readers who are not interested in the formalism of Floquet theory. This 
formalism, the \emph{Floquet picture}, will then be introduced in section 
\ref{sec:FloquetPicture} and employed to describe various effects beyond the simple 
approximations used in the preceding section \ref{sec:experiments}. Here we will discuss 
issues like those mentioned in the previous paragraph: heating, the asymptotic behavior in the 
long-time limit, anomalous topological edge states, and the effective adiabatic dynamics 
required for state preparation. We will close with conclusion and outlook in section
\ref{sec:conclusions}. 

The material and the references covered in the present article are selected as follows. We try 
to give a rather complete overview of the recent experiments, where periodic forcing was used 
to coherently control atomic quantum gases in optical lattices (not including experiments 
where modulation was employed for spectroscopic purposes). This includes the corresponding 
theoretical proposals and analyses. The theory of periodically driven many-body quantum 
systems recently became a very active field. We do not attempt to (and cannot) give an 
exhaustive overview of this rapidly growing field, but mention some pioneering contributions 
relevant for future quantum-gas experiments. This selection of covered works reflects the 
interests and the background of the author and is constrained by the format of a short review. 
It unavoidably misses contributions that would have been worth being covered as well. Further 
information and references about the Floquet theory of periodically driven quantum systems can 
be found in excellent recent review articles, covering the control of tunneling
\cite{GrifoniHaenggi98}, multiphoton processes in atoms and molecules \cite{ChuTelnov03}, 
AC-driven transport in nano-structured devices \cite{PlateroAguadoEtAl04, KohlerEtAl05},
high-frequency approximations \cite{GoldmanDalibard14, BukovEtAl15} and band-structure 
engineering \cite{Holthaus16}.

\section{\label{sec:GenProp}Some general properties of Floquet systems}
Let us consider quantum systems described by a time-periodic Hamiltonian 
\be
\Ho(t)=\Ho(t+T)=\sum_{m=-\infty}^\infty e^{im\omega t}\Ho_m,
\ee
with $\Ho_m \equiv \frac{1}{T}\int_0^T\!\rd t\,e^{-im\omega t}\Ho(t)=\Ho_{-m}^\dag$. The time 
evolution operator $\Uo(t,t_0)$ describes solutions $|\psi(t)\ra=\Uo(t,t_0)|\psi(t_0)\ra$ of 
the time-dependent Schr\"odinger equation 
\be\label{eq:schroedinger}
i\hbar\rd_t |\psi(t)\ra = \Ho(t)|\psi(t)\ra. 
\ee
At least formally, one can now construct a time-periodic unitary operator $U_F(t)=\Uo_F(t+T)$, 
such that the time evolution of the transformed state $|\psi_F(t)\ra=\Uo_F^\dag(t)|\psi(t)\ra$ 
is governed by a \emph{time-independent} Hamiltonian\footnote{Such a transformation might not 
exist in the limit of an infinite-dimensional states space \cite{GesztesyMitter81}. However, 
here we are dealing with systems of finite spatial extent on finite time scales, on which 
states above some high-energy cutoff will not matter.}
\be\label{eq:HFgauge}
\Ho_F = \Uo_F^\dag(t)\Ho(t)\Uo_F(t)-i\hbar\Uo_F^\dag(t)\dot{\hat{U}}_F(t).
\ee
In terms of these operators, the time evolution operator takes the form \cite{Shirley65}
\be\label{eq:tri}
\Uo(t,t_0)\equiv\Uo_F(t)\exp\Big({-\frac{i}{\hbar}(t-t_0)\Ho_F}\Big)\Uo_F^\dag(t_0).
\ee
It illustrates that the evolution of a Floquet system results from the interplay of two 
ingredients. On the one hand, the \emph{micromotion operator} $\Uo_F(t)$ describes a
time-periodic component of the dynamics, the micromotion. It can be expressed like
$\Uo_F(t)=e^{-i\hat{K}(t)}$ in terms of a hermitian time-periodic \emph{kick operator}
$\hat{K}(t)$ describing the effect of an abrupt switching on of the forcing
\cite{GoldmanDalibard14}. On the other hand, the time-independent \emph{effective Hamiltonian} 
$\Ho_F$ describes a linear phase evolution, which determines the time evolution in a similar 
way as a time-independent Hamiltonian determines the time evolution of an autonomous system.

The eigenvalue problem of the effective Hamiltonian
\be
\Ho_F|\tilde{u}_n\ra=\varepsilon_n|\tilde{u}_n\ra
\ee
gives rise to generalized stationary states $|\psi_n(t)\ra$ of the time dependent Schr\"odinger 
equation, called \emph{Floquet states}. They are of the form \cite{AutlerTownes55, Shirley65, 
Zeldovich67}
\be
|\psi_n(t)\ra=e^{-\frac{i}{\hbar}t\varepsilon_n}|u_n(t)\ra , 
\quad
|u_n(t)\ra=\Uo_F(t)|\tilde{u}_n\ra.
\ee
Here the periodic time-dependence of the \emph{Floquet mode} $|u_n(t)\ra=|u_n(t+T)\ra$ 
represents the micromotion and the quasienergy $\varepsilon_n$ determines the linear 
phase evolution. The Floquet states are eigenstates of the time-evolution operator over one 
driving period, 
\be\label{eq:Ueigen}
|\psi_n(t+T)\ra=\Uo(t+T,t)|\psi_n(t)\ra=e^{-\frac{i}{\hbar}T\varepsilon_n}|\psi_n(t)\ra.
\ee 
For every time $t$, they form a complete and orthogonal basis. If the system is prepared in a 
Floquet state, its time evolution is periodic, determined by the Floquet mode $|u_n(t)\ra$, and 
in this sense quasi stationary. If the system is prepared in a coherent superposition of 
several Floquet states, 
\be\label{eq:evolution}
|\psi(t)\ra=\sum_n c_n e^{-\frac{i}{\hbar}t\varepsilon_n}|u_n(t)\ra ,
\quad c_n=e^{\frac{i}{\hbar}t_0\varepsilon_n}\la u_n (t_0)|\psi(t_0)\ra,
\ee
deviations from a periodic evolution are governed by the quasienergies $\varepsilon_n$.  

There is not a unique time-periodic micromotion operator $\Uo_F(t)$ leading to a unique
time-independent effective Hamiltonian $\Ho_F$. Starting from one solution, $\Uo_F(t)$, another
one, $\Uo_F'(t)$, can be constructed by applying certain  operations. The Floquet states
$|\psi_n(t)\ra$, being eigenstates of the time-evolution operator, will not be altered by such 
operations. The simplest possibility is to multiply the micromotion operator with an arbitrary 
time-independent unitary operator $\Uo$ from the right, $\Uo_F'(t)=\Uo_F(t)\Uo$, so that
$\Ho_F'=\Uo^\dag\Ho_F\Uo$. For example, by choosing
$\Uo_F'(t)=\Uo_F(t)\Uo_F^\dag(t_0)\equiv\Uo_F(t,t_0)$ a new micromotion operator is obtained 
that becomes equal to the identity once during each driving period,
$\Uo'_F(t_0)=\Uo_F(t_0,t_0)=1$. This allows for writing the time-evolution operator like 
\be\label{eq:bi}
\Uo(t,t_0)\equiv\Uo_F(t,t_0)\exp\Big({-\frac{i}{\hbar}(t-t_0)\Ho^F_{t_0}}\Big),
\ee
with \emph{Floquet Hamiltonian}
\be\label{eq:HfloquetHeff}
\Ho^F_{t_0}= \Uo_F(t_0)\Ho_F\Uo_F^\dag(t_0)
\ee
and \emph{two-point micromotion operator}
\be\label{eq:U2tU1t}
\Uo_F(t,t_0) = \Uo_F(t)\Uo_F^\dag(t_0).
\ee
In particular, for $t=t_0+T$ Eq.~(\ref{eq:bi}) reduces to Eq.~(\ref{eq:UHeff}). 
The Floquet Hamiltonian is a special choice of the effective Hamiltonian, which directly 
generates the stroboscopic time evolution in steps of the driving period $T$. $\Ho^F_{t_0}$ 
depends parametrically on the initial time $t_0$, and thus also on the driving phase. However, 
according to Eq.~(\ref{eq:HfloquetHeff}) this dependence is rooted in a unitary transformation, 
so that the spectrum of the Floquet Hamiltonian is independent of $t_0$ and the driving phase. 

Another possibility to construct a new micromotion operator and effective Hamiltonian is given 
by $\Uo_F'(t)=\Uo_F(t)\exp\big(im\omega t |\tilde{u}_n\ra\la\tilde{u}_n|\big)$ with integer $m$,
which implies $\Ho_F'=\Ho_F+m\hbar\omega|\tilde{u}_n\ra \la\tilde{u}_n|$. This operation 
changes the quasienergy $\varepsilon_n$ and its Floquet mode to new solutions labeled by $m$:
\be\label{eq:eps_nm}
\varepsilon_{nm}=\varepsilon_n+m\hbar\omega, 
\quad
|u_{nm}(t)\ra = e^{im\omega t}|u_n(t)\ra,
\ee
such that the corresponding Floquet state is not altered, 
\be\label{eq:psi_nm}
|\psi_n(t)\ra=e^{-\frac{i}{\hbar}t\varepsilon_n}|u_n(t)\ra
	=e^{-\frac{i}{\hbar}t\varepsilon_{nm}}|u_{nm}(t)\ra.
\ee
Equation~(\ref{eq:eps_nm}) shows that quasienergies are defined up to integer multiples of
$\hbar\omega$ only, in agreement with the earlier observation that Eq.~(\ref{eq:UHeff}) does 
not determine the Floquet Hamiltonian uniquely. This property reflects the possibility of 
resonant coupling. The freedom to choose $m$ individually for each Floquet state $n$, can be 
used to choose all quasienergies to lie in the same interval of width $\hbar\omega$. Such an 
interval is often called \emph{Brillouin zone}, in loose analogy to Bloch's theory of 
spatially periodic systems. In the latter quasimomenta are defined modulo reciprocal lattice 
vectors only. They can be chosen to lie in one elementary cell of the reciprocal lattice such 
as the first Brillouin zone. In case the Floquet mode
$|u_n(t)\ra\equiv \sum_{m'}|u_n^{(m')}\ra e^{-im'\omega t}$ is 
dominated by a specific harmonic $m'=m_0$ with respect to a given frame of reference, $m=m_0$ 
constitutes a meaningful choice for the quasienergy, which, in the limit of a time-independent 
Hamiltonian, reproduces the energy spectrum. 

A prerequisite for Floquet engineering is a theoretical method to compute the effective 
Hamiltonian and the micromotion operator, at least within a suitable approximation. For
$\hbar\omega$ large compared to the matrix elements of the Hamiltonian, a systematic 
approximation to the effective Hamiltonian and the micromotion operator is given by a
high-frequency expansion \cite{GrozdanovRakovic88, RahavEtAl03, GoldmanDalibard14, 
ItinKatsnelson15, GoldmanEtAl15, EckardtAnisimovas15, MikamiEtAl16})
\be\label{eq:hf}
\Ho_F \approx \sum_{\mu=1}^{\mu_\text{cut}}\Ho_F^{(\mu)},\qquad
\Uo_F(t)\approx \exp\bigg(\sum_{\mu=1}^{\mu_\text{cut}}\hat{G}^{(\mu)}(t)\bigg).
\ee
Here $\Ho_F^{(\mu)\dag}=\Ho_F^{(\mu)}$ and $[\hat{G}^{(\mu)}(t)]^\dag=-\hat{G}^{(\mu)}(t)$.
The leading terms are given by
\bes\label{eq:Hhf}
\Ho_F^{(1)}&=&\Ho_0,
\qquad 
\Ho_F^{(2)} = \sum_{m\ne0}\frac{\Ho_m\Ho_{-m}}{m\hbar\omega},
\nonumber\\
\Ho_F^{(3)} 
	&=& \sum_{m\ne0}\Bigg[\frac{\big[\Ho_{-m},\big[\Ho_0,\Ho_m\big]\big]}{2(m\hbar\omega)^2}
\nonumber\\&&
	+\,\sum_{m'\ne 0,m}\frac{\big[\Ho_{-m'},\big[\Ho_{m'-m},\Ho_m\big]\big]}
											{3mm'(\hbar\omega)^2}\Bigg],
\ees
and
\bes\label{eq:Ghf}
\hat{G}^{(1)}(t) &=& -\sum_{m\ne0}\frac{e^{im\omega t}\Ho_m }{m\hbar\omega},
\nonumber\\
\hat{G}^{(2)}(t)&=&\sum_{m\ne0}
	\Bigg[\frac{e^{im\omega t}\big[\Ho_0,\Ho_m\big]}{(m\hbar\omega)^2}
\nonumber\\&&
	+\,\sum_{m'\ne 0,m}\frac{e^{i(m-m')\omega t}\big[\Ho_{-m'},\Ho_m\big]}
									{2m(m-m')(\hbar\omega)^2}\Bigg].
\ees

A similar high-frequency expansion for the Floquet Hamiltonian $\Ho^F_{t_0}$ and the two-point 
micromotion operator $\Uo_F(t,t_0)$ is known as Floquet-Magnus expansion
\cite{Maricq82,MilfeldWyatt83,CasasEtAl00,BlanesEtAl09,BukovEtAl15,VerdenyEtAl13}
\be\label{eq:FM}
\Ho_{t_0}^F \approx \sum_{\mu=1}^{\mu_\text{cut}}\Ho_{t_0}^{F(\mu)},\qquad
\Uo_F(t,t_0)\approx \exp\bigg(\sum_{\mu=1}^{\mu_\text{cut}}\hat{F}^{(\mu)}(t,t_0)\bigg).
\ee
We can contruct the leading terms from the expansion (\ref{eq:hf}) using
Eqs.~(\ref{eq:HfloquetHeff}) and (\ref{eq:U2tU1t}). For the Floquet-Hamiltonian they read
\bes\label{eq:H_FM}
\Ho_{t_0}^{F(1)}&=&\Ho_F^{(1)}=\Ho_0,
\nonumber\\
\Ho_{t_0}^{F(2)} &=& \Ho_F^{(2)} - [\Ho_F^{(1)},G^{(1)}(t_0)]
\nonumber\\&=& 
\sum_{m\ne0}\frac{\Ho_m\Ho_{-m}+e^{im\omega t_0}[H_0,H_m] }{m\hbar\omega}.
\ees
Here the second term of $\Ho_{t_0}^{F(2)}$ results from the expansion of the unitary operator
$\Uo_F(t)\simeq 1 + \hat{G}^{(1)}(t)+\cdots$. This expansion conserves unitarity only up to the 
considered order $\mu_\text{cut}$, e.g.\ in first order one finds
$[1 + \hat{G}^{(1)}(t)]^\dag[1 + \hat{G}^{(1)}(t)]=1-[\hat{G}^{(1)}(t)]^2$. As a consequence, 
the approximate quasienergy spectrum obtained from the Floquet Hamiltonian in
$\mu_\text{cut}$th-order acquires a spurious dependence on the initial time $t_0$ and, thus, 
also on the driving phase. When expanding the spectrum in powers of the inverse driving 
frequency, the $t_0$ dependence appears in terms of powers $\ge \mu_\text{cut}$, which cannot 
be expected to be captured correctly within the given order of the approximation. While these 
terms should be small in the regime where the approximation is justified, they might cause 
spurious symmetry breaking \cite{EckardtAnisimovas15}.

The Floquet-Magnus expansion is guaranteed to converge, if the period-averaged operator norm 
of the Hamiltonian $\Ho(t)$ is smaller than $\xi_F\hbar\omega$,  where $\xi_F$ is a constant 
of order one \cite{CasasEtAl00}. For periodically driven many-body systems, possessing excited 
states also at macroscopically large energies, this condition cannot be expected to be 
fulfilled (unless the state space is effectively reduced by symmetry or localization). 
However, even in this case the high-frequency expansion might still provide a suitable 
approximation provided $\hbar\omega$ is large compared to the typical intensive energy scales 
of the system, at least up to a certain time span $t_h$ beyond which the system heats up
\cite{Maricq82, AbaninEtAl16, KuwaharaEtAl16, MoriEtAl16}. 
\onlinecite{KuwaharaEtAl16, AbaninEtAl16} showed that for spin systems with local interactions 
(i.e.\ for systems with local energy bound) the time scale $t_h$ increases exponentially with 
the driving frequency. For these systems, \onlinecite{KuwaharaEtAl16}, moreover, showed that 
for time spans smaller than $t_h$ the Floquet-Magnus expansion is (at least) an asymptotic 
series that provides a good approximation for the time-evolution operator, whose error rapidly 
decreases with $\mu_\text{cut}$ before it increases again beyond an optimal order
$\mu_\text{cut}^\text{opt}$. While these results do not apply to optical lattice systems,
which do not have a local energy bound, they still indicate that the high-frequency 
approximation can provide an accurate description of a driven many-body system as long as the 
duration of the experiment is short compared to some heating time $t_h$. This issue is 
discussed in more detail in section~\ref{sec:FloquetPicture}.

The approximate effective Hamiltonian, as it is given by a certain low order $\mu_\text{cut}$ 
of the high-frequency approximation (\ref{eq:hf}), defines a simple model Hamiltonian. In 
contrast, the full effective Hamiltonian of a driven system of many interacting particles is 
typically a highly complex (rather awkward) object, which cannot be written down explicitly. 
Very often the starting point of Floquet engineering is, therefore, to realize the physics of 
an autonomous model described by a target Hamiltonian $\Ho_\text{target}$ directly 
corresponding to the high-frequency approximation in some low order ($\mu_\text{cut}=1$ or $2$),
$\Ho_\text{target}=\sum_{\mu=1}^{\mu_\text{cut}}\Ho_F^{(\mu)}$.
From this perspective, the dynamics of the driven quantum system provides an approximation to 
the physics of the desired model Hamiltonian $\Ho_\text{target}$, rather than the other way 
around. The quantum-gas experiments described in the following section \ref{sec:experiments} 
can be interpreted from this point of view.

\section{\label{sec:experiments}Quantum-gas experiments and their basic description}

This section shall give an overview over recent experiments with quantum gases of ultracold 
neutral atoms in periodically driven optical lattices. We will not discuss experiments, where 
periodic driving has been employed for spectroscopic purposes, but rather describe those 
aiming for the coherent manipulation of the system's state and its dynamics. The observed 
effects will be explained in terms of a common language and using intuitive approximations.

\subsection{Neutral atoms in optical lattices}

Ultracold quantum gases \cite{BlochDalibardZwerger08, LewensteinSanperaAhufinger} consist of 
neutral atoms held in optical or magneto-optical traps inside a vacuum cell and cooled down to 
quantum degeneracy by means of laser cooling and evaporative cooling. They are very well 
isolated from their environment. Dissipative processes, such as the formation of molecules via 
three-body collisions, spontaneous emission as a result of the optical trap, or collisions with 
background particles, are often negligible. Atom numbers of up to several millions can be 
reached. 

The possibility to create light-shift potentials proportional to the laser intensity allows for 
the creation of quasi defect-free lattice potentials from standing light waves, called
\emph{optical lattices}. For example, a one-dimensional lattice created by two 
counter-propagating laser beams with wave vectors $\bk_L$ and $-\bk_L$ takes the form of a 
cosine lattice, $V_L(\br)=V_0\sin^2(\bk_L\cdot\br)$, where the lattice depth $V_0$ 
is proportional to the laser intensity. Besides the lattice depth $V_0$, a second energy scale 
is the recoil energy $E_R=\hbar^2k_L^2/2m$, with $k_L=|\bk_L|$ and atomic mass $m$. It 
corresponds to the kinetic energy required to localize a particle on the length of a lattice 
constant $d=\pi/k_L$. Recoil energies are of the order of several ($h\times$) kilohertz, 
roughly corresponding to ($k_B\times$) microkelvin or several pico-electron volts. The lattice 
depth can take values of up to hundreds of recoil energies. The rather large time scales 
corresponding to these low energy scales allow for accurate time-dependent manipulation and 
time-resolved imaging. By combining standing waves in different direction or by creating more 
complex interference patterns one can create various two- and three-dimensional lattice 
structures. Moreover, effectively one- or two-dimensional systems can be realized by strong 
transversal confinement. 

For deep lattices, $V_0\gg E_R$, the system is well described by a Hubbard model with one 
localized Wannier state at each lattice minimum. The single-particle terms of the Hamiltonian 
take the form
\be\label{eq:Hsp}
\Ho_\text{tun} =  -\sum_{\la\ell'\ell\ra}J\aa_{\ell'}\ao_\ell,
\qquad \Ho_\text{pot} = \sum_\ell v_\ell \no_\ell 
\ee
where $\aa_\ell$, $\ao_\ell$, and $\no_\ell=\aa_\ell\ao_\ell$ denote the creation, annihilation 
and number operator for a particle, boson or fermion, in the Wannier state at lattice site
$\ell$. The kinetics is captured by $\Ho_\text{tun}$ and to good approximation exhausted by 
tunneling processes between neighboring sites $\ell$ and $\ell'$, here $\la\ell'\ell\ra$ 
denotes a directed pair of neighboring sites $\ell'$ and $\ell$. The nearest-neighbor tunneling 
parameter $J$ depends sensitively on the lattice depth and can, depending on the lattice 
structure, also acquire a directional dependence $J\to J_{\ell'\ell}$. 
For a deep cosine lattice $J/E_R\simeq 4\pi^{-1/2}(V_0/E_R)^{3/4}\exp(-2\sqrt{V_0/E_r})$
\cite{Zwerger03}. The potential term $\Ho_\text{pot}$ captures the influence of an external 
potential such as the trap or a superlattice potential. 

The interactions among low-temperature alkaline atoms, as they were used in the experiments to 
be reviewed here, are short-ranged and captured by on-site terms. For spinless bosons the 
interaction term reads
\be\label{eq:Hint}
\Ho_\text{int} =\frac{U}{2}\sum_\ell\no_\ell(\no_\ell-1).
\ee 
For deep lattices the Hubbard parameter $U$ approaches
$U\simeq\sqrt{2/\pi}\hbar^2a_s/(m\bar{a}_0^3) =2E_R\sqrt{2/\pi}(k_L a_s)/(k_L\bar{a}_0)^3$ with
s-wave scattering length $a_s$ and the mean harmonic-oscillators length $\bar{a}_0$ in the 
lattice minimum. For the cosine lattice the harmonic-oscillator length depends weakly on the 
lattice depth like $a_0k_L=(V_0/E_R)^{-1/4}$ and for Rb-87 atoms $k_L a_s\approx0.041$ at 
$2\pi/k_L=850$ nm. Spinless (i.e.\ spin-polarized) fermions do not interact due to Pauli 
exclusion. In order to have interactions among fermionic atoms, one has to consider spinful 
atoms or elements with long-ranged dipolar interactions.

The Hubbard model is justified for sufficiently deep lattices ($V_0/E_R>5$) and has been tested 
to provide a quantitative description of optical lattice systems \cite{TrotzkyEtAl09}. Excited 
states belonging to higher Bloch bands not included in the Hubbard model are separated by a 
large energy gap $E_G$ of several $E_R$. The gap $E_G$, which is roughly given by 
$E_G\approx E_R\min\big(V_0/(2E_R),2\sqrt{V_0/E_R}-1\big)$, can be two orders of magnitude 
larger than $J$ and $U$. Thus, even if the driving frequency is required to be large compared 
to $J$ and $U$, it can still be small compared to the band gap. This suggests that a 
description of the periodically driven systems in terms of the low-energy subspace described 
by the Hubbard model is possible. A more detailed discussion of this issue is given in
Sec.~\ref{sec:heating} below. 

\subsection{\label{sec:dl}Dynamic localization}

The first experiment where the coherent dynamics of an ultracold quantum gas has been 
controlled by means of periodic forcing has been conducted in Arimondo's group in Pisa. The 
ballistic spreading of a localized Bose-Einstein condensate in the lowest band of a
one-dimensional optical lattice has been slowed down, and even suppressed completely, by the 
application of a sinusoidal force (\onlinecite{LignierEtAl07}, see also
\onlinecite{EckardtEtAl09,CreffieldEtAl10}). This is the effect of \emph{dynamic localization}
\cite{DunlapKenkre86}. 

The experimentalists created a one-dimensional optical lattice in the tight-binding regime 
along the $x$ direction, together with a tube-like harmonic confinement in the radial directions $y$
and $z$. Initially a Bose-Einstein condensate of $^{87}$Rb atoms was loaded into the lowest 
Bloch band of the lattice, localized in the center of the tube by an additional trapping 
potential. When this additional trap was switched off, the condensate started to expand 
in the tube. During this expansion a sinusoidal force was applied, created as an inertial 
force by shaking the lattice back and forth. The shaken lattice is described by the potential
$V_{DL}(\br,t)=V_L(\br-\bxi(t))$ with $\bxi(t)=\bxi(t+T)=\bxi(t)=\xi_0\cos(\omega t){\bm e}_x$, 
which transforms to $V_L(\br)-\br\cdot\bF(t)$ in the reference frame co-moving with the 
lattice, where 
\be\label{eq:Fin}
\bF(t)=-m\ddot{\bxi}(t)=\bF_\omega\cos(\omega t)
\ee 
with $\bF_\omega=m\omega^2\xi_0{\bm e}_x\equiv F_\omega {\bm e}_x$ \cite{DreseHolthaus97,
MadisonEtAl98}. After a certain time of expansion in the driven lattice, the atom density was 
measured by absorption imaging either \emph{in situ} or, in order to reveal the momentum 
distribution, after an additional time of flight with all potentials switched off. 

\begin{figure}[t]
\includegraphics[width=0.6\linewidth]{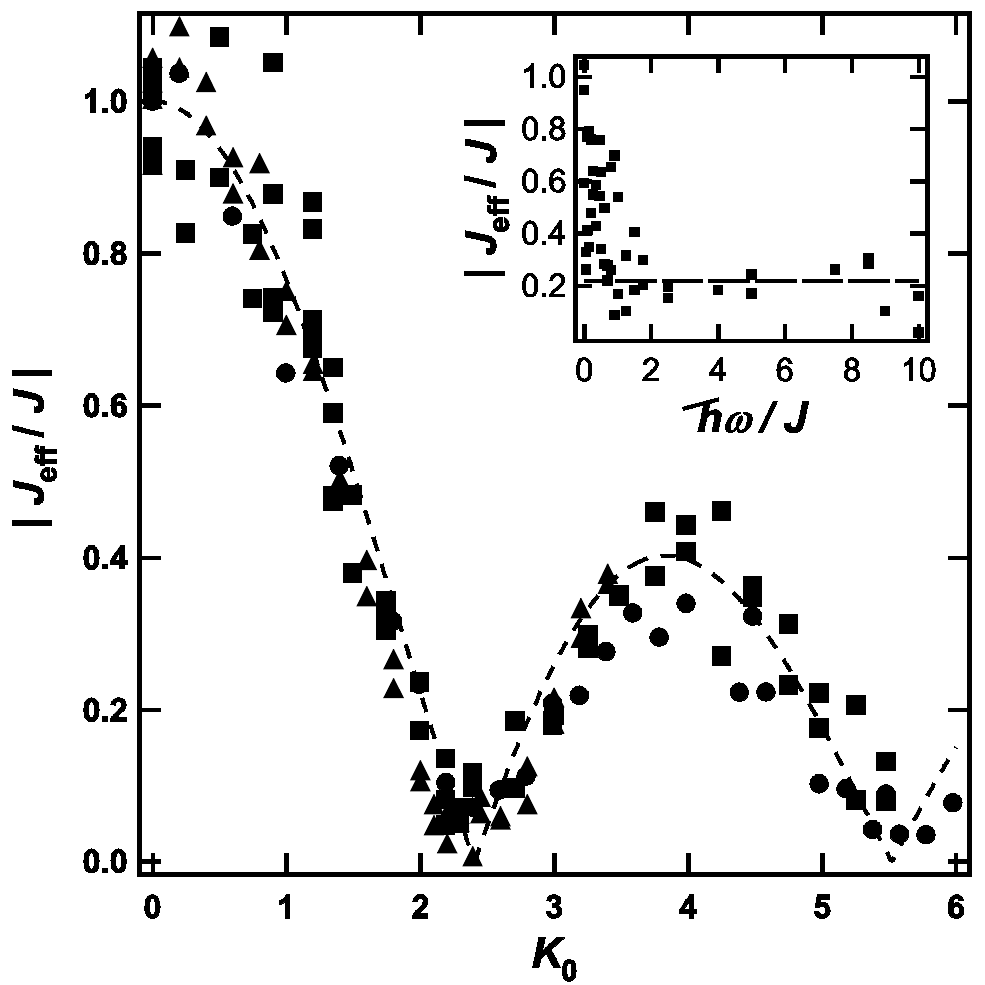}
\centering
\caption{\label{fig:DL} 
Effective tunneling matrix element $|J_\text{eff}|/J$ versus $K_0=K/\hbar\omega$. Extracted from the expansion dynamics 
of a condensate of about $5\cdot 10^4$ $^{87}$Rb atoms in a shaken optical lattice of $E_R\approx 2\pi\hbar 3.16 $ kHz
(squares: $V_0/E_R=6$, $\omega/2\pi=1$ kHz, circles: $V_0/E_R=6$, $\omega/2\pi=0.5$ 
kHz, triangles: $V_0/E_R=4$,$\omega/2\pi=1$ kHz, dashed line: theoretical prediction). Inset: 
$J_\text{eff}/J$ for $K/\hbar\omega=2$ and $V_0/E_R=9$ versus $\hbar\omega/J$, indicates breakdown of
high-frequency prediction (dashed line) for $\hbar\omega/J<2$. (taken from \onlinecite{LignierEtAl07})} 
\end{figure}

For a broad range of parameters, the momentum distribution revealed sharp peaks, indicating 
that the condensate retained its coherence like in the case of a ballistic expansion and that 
the shaking did not cause significant heating. Moreover, by comparing the \emph{in-situ} extent 
of the atom cloud with that found for ballistic expansion in the non-driven lattice
(Fig.~\ref{fig:DL}), the driven system was found to be well-described by the effective 
tunneling parameter predicted by \onlinecite{DunlapKenkre86}, 
\be\label{eq:Jeff}
J_\text{eff}=J \Jb_0\Big(\frac{K}{\hbar\omega}\Big),
\ee
where $K=dF_\omega$ is the amplitude of the potential modulation between neighboring lattice 
sites.

Let us explain this result. In the lattice frame of reference the system can be described by 
the tight-binding Hamiltonian 
\be\label{eq:DBH}
\Ho(t)= -\sum_{\la\ell'\ell\ra} J \aa_{\ell'}\ao_\ell 
	+ \sum_\ell\bigg[\big(v_{\ell}+w_{\ell}(t)\big)\no_\ell 
+\frac{U}{2}\no_\ell(\no_\ell-1)\bigg],
\ee
where $\ell$ labels the minima $\br_\ell=(x_\ell,0,0)$ of the one-dimensional lattice and where
\be
w_\ell(t)=-\br_\ell\cdot \bF(t)= -\frac{x_\ell}{d} K\cos(\omega t),
\qquad v_\ell=v^\text{tr}_\ell
\ee
captures the periodic force as well as a weak harmonic trap in the lattice direction
$v^\text{tr}_\ell$, respectively. In the expansion experiment $v_\ell^\text{tr}$ was small 
enough to have no significant influence on the measured expansion dynamics.  

One can now perform a gauge transformation, 
$|\psi'(t)\ra=\Uo^\dag(t)|\psi(t)\ra$ and $\Ho'(t) 
= \Uo^\dag(t)\Ho(t)\Uo(t)-i\hbar\Uo^\dag(t)[\rd_t\Uo(t)]$,
defined by the time-periodic unitary operator  
\be\label{eq:Uchi}
\Uo(t)=\exp\bigg(i\sum_\ell \chi_\ell(t)\no_\ell \bigg)
\ee
with
\be\label{eq:chi}
\chi_\ell(t)=-\int_{t_0}^t\!\rd t' \frac{w_\ell(t')}{\hbar} 
		-\chi_{0\ell}. 
\ee
The time-independent gauge constant $\chi_{0\ell}$ shall be chosen such that
$\int_0^T\!\rd t\,\chi_\ell(t)=0$. 
Writing 
\be
\chi_\ell(t)\equiv\br_\ell\cdot \ba(t)
\ee
reveals that the unitary operator $\Uo(t)$ describes a global shift in quasimomentum by
\be
\ba(t)=-\frac{m}{\hbar}\dot{\bxi}(t) 
= \frac{1}{d} \frac{K}{\hbar\omega}\sin(\omega t){\bm e}_x .
\ee
By employing $\Uo^\dag(t)\ao_\ell\Uo(t)=e^{i\chi_\ell(t)}\ao_\ell$ and noting that the
time-derivative of the unitary transformation cancels with the driving term, the
gauge-transformed Hamiltonian can be brought to the form 
\be\label{eq:Hprime}
\Ho'(t)=-\sum_{\la\ell'\ell\ra} J e^{i\theta_{\ell'\ell}(t)}\aa_{\ell'}\ao_\ell 
		+\sum_\ell\bigg[v_\ell\no_\ell+\frac{U}{2}\no_\ell(\no_\ell-1)\bigg].
\ee
The time-dependent Peierls phases 
\be\label{eq:Peierls}
\theta_{\ell'\ell}(t)=\chi_\ell(t)-\chi_{\ell'}(t) = -(\br_{\ell'}-\br_{\ell})\cdot \ba(t)
\ee
play the role of a discrete vector potential that now, instead of the discrete scalar 
potential $w_\ell(t)$, describes the force $\bF(t)$. 

For $v_\ell=0$, the gauge transform restores the discrete translational symmetry of the 
lattice. 
The time-dependent tunneling term of $\Ho'(t)$ is, thus, diagonal in quasimomentum 
representation, 
\be\label{eq:H'tun}
\Ho'_\text{tun}(t) = \sum_{\bk} \varepsilon\big(\bk+\ba(t)\big)\no_\bk.
\ee
Here $\no_\bk$ denotes the number operator for particles in the Bloch state with quasimomentum 
wave number $\bk$, characterized by $\la\ell|\bk\ra\propto\exp(i\br_\ell\cdot\bk)$, and
$\varepsilon(\bk)=-2J\cos(dk_x)$ is the single-particle dispersion relation of the undriven 
tight-binding lattice. The form (\ref{eq:H'tun}) explicitly shows that the transformation
(\ref{eq:Uchi}) describes a shift in momentum by $\ba(t)$.\footnote{The transformation
can be viewed as an analog of the Kramers-Henneberger transformation \cite{Henneberger68}, 
with the roles of momentum and position interchanged.} 
Under the influence of $\Ho_\text{tun}'(t)$ the quasimomentum occupation 
numbers of the state do not change in time. During each driving period a particle in state
$|\bk\ra$ just pics up an integrated dynamical phase proportional to the time-averaged energy
$\varepsilon_\text{eff}(\bk)=\frac{1}{T}\int_0^T\!\rd t\,\varepsilon\big(\bk+\ba(t)\big)
=-2J_\text{eff}\cos(dk_x)$, with the effective tunneling matrix element $J_\text{eff}$ as given 
by Eq.~(\ref{eq:Jeff}).\footnote{In order to compute the time average, the identity
$\exp(i a\sin(b)) = \sum_{\mu=-\infty}^\infty \Jb_\mu(a)\exp(i\mu b)$ was employed.} Thus, 
apart from an oscillatory dynamics at the driving frequency the system behaves as if it was 
described by the effective dispersion relation $\varepsilon_\text{eff}(\bk)$ with a reduced 
band width of $4J_\text{eff}$. This argument is valid also for low driving frequencies, unless 
interactions or translational-symmetry-breaking cause scattering between quasimomentum states. 

In the more interesting situation with interactions and translational-symmetry breaking, the 
effective modification of tunneling is still valid approximately in the high-frequency regime. 
Namely, if $\hbar\omega$ is large compared to the characteristic energy scales $J$, $U$, and 
$|v_{\ell'}-v_\ell|$ on neighboring sites $\ell'$ and $\ell$, which determine the rates at 
which the system's state $|\psi'(t)\ra$ changes in time, we can average over the rapid 
oscillation of the Peierls phases. The tight-binding Hamiltonian can be approximated by its 
cycle average 
\be\label{eq:RW}
\Ho'(t)\approx\frac{1}{T}\int_0^T\!\rd t\Ho'(t)\equiv\Ho_\text{eff}.
\ee
One finds
\be\label{eq:Heff}
\Ho_\text{eff}=-\sum_{\la\ell'\ell\ra} J^\text{eff}_{\ell'\ell} \aa_{\ell'}\ao_\ell 
		+\sum_\ell\bigg[v_\ell \no_\ell+\frac{U}{2}\no_\ell(\no_\ell-1)\bigg],
\ee
with modified tunneling matrix element
\be\label{eq:Jeff_ell}
J^\text{eff}_{\ell'\ell} = \frac{J}{T}\int_0^T \!\rd t\, e^{i[\chi_\ell(t)-\chi_{\ell'}(t)]},
\ee
resulting in $J^\text{eff}_{\ell'\ell}=J_\text{eff}$, where $J_\text{eff}$ is given by
Eq.~(\ref{eq:Jeff}).\footnote{For non-sinusoidal square-wave forcing the modification of 
tunneling is given by a sinc function \cite{ZhuEtAl99}, as it was observed also
experimentally \cite{EckardtEtAl09}.} This rotating-wave-type approximation is in principle 
valid not only for the case of weak interactions of the expansion experiment, but also in the 
regime where $U$ is comparable or larger than $J$ \cite{EckardtEtAl05b}.

The result of the rotating-wave approximation can be related to Floquet theory. The
time-independent Hamiltonian $\Ho_\text{eff}$ that was argued to effectively describe the time 
evolution, constitutes an approximation to the effective Hamiltonian $\Ho_F$ and the unitary 
operator (\ref{eq:Uchi}) approximates the micromotion operator $\Uo_F(t)$: 
\be\label{eq:RW}
\Ho_F\approx \Ho_\text{eff},\qquad \Uo_F(t)\approx \Uo(t).
\ee
This corresponds to the leading order of the high-frequency approximation (\ref{eq:hf}) applied 
to $\Ho'(t)$, 
\be
\Ho_\text{eff}=\Ho_F^{'(1)}.
\ee

As will be discussed in section~\ref{sec:FloquetPicture} this approximation is expected to be 
valid on a certain time scale, before heating sets in. This time scale can, however, be rather 
long and comparable to the duration of the experiment (typically a few hundred milliseconds). 
In the experiment by \onlinecite{LignierEtAl07}, the condensate coherence was found to decay 
during the expansion on a dephasing time of about 200 milliseconds for the non-driven system 
with $V_0/E_R\approx9$. For the strongly driven system with $K/\hbar\omega=2.2$ comparable 
dephasing times were achieved when the driving frequency was increased to an optimal value.

A significant modification of tunneling requires strong driving with the driving strength $K$ 
of the order of $\hbar\omega$. For such strong forcing it would not have been justified to 
approximate the original Hamiltonian $\Ho(t)$ by its time average, since the amplitude of the 
driving term $K$ changes the state at a rate comparable to the driving frequency. However, by 
integrating out the driving term via a gauge transformation before applying the rotating-wave 
approximation, also the case of strong driving can be treated \cite{EckardtEtAl10, 
GoldmanEtAl15}. In this way a non-perturbative treatment of the forcing has been achieved. 
This is visible in the fact that through the Bessel-function-type dependence the effective 
tunneling matrix element $J_\text{eff}$ contains arbitrarily large powers of the driving 
amplitude $K$. In contrast, when performing the high-frequency expansion starting from $\Ho(t)$
in $\mu$th order the largest power encountered is $K^\mu$. Thus, integrating out the driving 
term via the gauge transformation (\ref{eq:Uchi}) before employing the high-frequency 
expansion corresponds to a partial resummation of the series (\ref{eq:hf}).

\onlinecite{LignierEtAl07} also observed deviations from the tight-binding description
(\ref{eq:DBH}) in the form of a small amount of transfer to the first excited band of the 
lattice (less than ten percent for $K/\hbar\omega<3$). Moreover, for negative effective 
tunneling matrix elements $J_\text{eff}<0$, the measured momentum distribution revealed that 
the atoms recondensed into the minimum $k_x=\pi/d$ of the inverted dispersion relation
$\varepsilon_\text{eff}(\bk)$. A plausible explanation of this process, during which the
(effective) kinetic energy is lowered, is that the excess energy is absorbed by excitations in 
the rather weakly confined transversal direction. 

The fact that the periodic force has been created as an inertial force via lattice shaking has 
a convenient implication concerning the measurement of the quasimomentum distribution. In the 
lattice frame of reference the quasimomentum distribution oscillates like $-\ba(t)$ in response 
to the inertial force. This oscillation is removed, when transforming back to the laboratory 
frame of reference, where the quasimomentum distribution is measured. This resembles the 
effect of the gauge transformation that led to $\Ho'(t)\approx\Ho_\text{eff}$. Time-of-flight 
pictures, thus, directly reveal the quasiomentum distribution of $\Ho_\text{eff}$ (multiplied, 
however, by an oscillating envelop given by the Fourier transform of the Wannier function).

\subsection{``Photon''-assisted coherent tunneling}

The starting point of a second type of experiment has been the Wannier-Stark configuration, 
namely a one-dimensional lattice system in combination with a homogeneous static force
$\bF_0=F_0{\bm e}_x$. If the potential difference between neighboring sites
$\Delta = F_\text{0}d$ is large compared to the band width (while still being small with 
respect to the band gap) tunneling processes between neighboring lattice sites are strongly 
suppressed. In this regime the localized single-particle Wannier-Stark eigenstates are 
approximately identical to the Wannier states at the lattice sites $\ell$, and Bloch 
oscillations are reduced to a rapid shivering motion of angular frequency $\Delta/\hbar$ and 
negligible amplitude $\sim J/\Delta$. An initially localized Bose condensate does not spread 
in time. However, coherent tunneling can be induced by applying a time-periodic force
$\bF_\omega\cos(\omega t)$, provided the resonance condition 
\be\label{eq:Delta}
\Delta=\nu\hbar\omega+\delta,
\ee
with integer $\nu$ and small detuning $\delta$ is met. This phenomenon is known as ``photon''-
assisted, AC-induced, or laser-assisted tunneling. For such a situation the energy separation
$\Delta$ between neighboring sites is bridged by $\nu$ energy quanta $\hbar\omega$ (tunneling 
corresponds to an allowed $\nu$-``photon'' transition) and particles can tunnel with an 
effective tunneling matrix element $-J_\text{eff}$, as if there was no potential tilt
\cite{Zak93}. If the resonance condition is not met exactly such that a finite detuning
$\delta$ remains, this detuning plays the role of a residual static force
$\delta \bF = (\delta/d){\bm e}_x$ \cite{EckardtHolthaus07}. 



A basic theoretical description of ``photon''-assisted tunneling in a sinusoidally forced 
tilted lattice can be obtained in a similar way as for the phenomenon of dynamic localization. 
Subjecting a one-dimensional lattice system to the force 
\be
\bF(t)=-\bF_0+\bF_\omega\cos(\omega t),
\ee
with $\bF_0=F_0{\bm e}_x$ and $\bF_\omega=F_\omega{\bm e}_x$, it is described by the
Bose-Hubbard Hamiltonian (\ref{eq:DBH}). However now the driving potential is defined like
\be\label{eq:vWS}
w_\ell(t)= \frac{x_\ell}{d}\big[-K\cos(\omega t)+\nu\hbar\omega\big],
\ee
where $K=F_\omega d$ as before, and 
\be\label{eq:well}
v_\ell = v^\text{tr}_\ell + \frac{x_\ell}{d}\delta,
\ee
describes an additional weak static potential. Here we have included the larger share
$\nu\hbar\omega$ of the static potential tilt (\ref{eq:Delta}) to the \emph{driving term}
$w_\ell(t)$, whereas the small detuning $\delta$ was included into $v_\ell$.

Now $w_\ell(t)$ contains all terms of the Hamiltonian, whose characteristic 
energy scale is not small compared to the driving frequency. Moreover the $w_\ell(t)$ term is 
defined such that it can be integrated out by a time-periodic gauge transformation, described 
by the unitary operator (\ref{eq:Uchi}) and (\ref{eq:chi}) with $w_\ell(t)$ as defined by
Eq.~(\ref{eq:vWS}). This gauge transformation leads to a Hamiltonian of the form
(\ref{eq:Hprime}). It corresponds to a global shift in quasimomentum by  
\be\label{eq:aWS}
\ba(t)=\Big[\frac{1}{d}\frac{K}{\hbar\omega}\sin(\omega t)-\frac{1}{d}\nu\omega t 
			+a_0\Big]{\bm e}_x,
\ee
where the constant $a_0$ depends both on the integration time $t_0$ and the gauge constant
$\chi_{0\ell}$ in Eq.~(\ref{eq:chi}). The linear dependence on time makes the definition of
$\chi_{0\ell}$ as the time average of the integral in Eq.~(\ref{eq:chi}) meaningless. Instead 
the freedom to choose $\chi_{0\ell}$ can be used either to incorporate into $a_0$ the actual 
momentum shift induced when both $\bF_\omega$ and $\bF_0$ are switched on according to a 
particular experimental protocol \cite{CreffieldSols11} or to achieve $a_0=0$ for convenience, 
as it shall be done in the following. The quasimomentum shift (\ref{eq:aWS}) is time periodic, 
in the sense that quasimomentum wave numbers $k_x$ are defined modulo $\frac{2\pi}{d}$ only. 
The integer $\nu$ corresponds to the number of times the system is translated in quasimomentum 
through the first Brillouin zone. 

We are now again in the position to approximate $\Ho'(t)=\Ho'(t+T)$ by its time average, as 
long as $J$, $U$, and $|v_{\ell'}-v_\ell|$ on neighboring sites $\ell'$ and $\ell$ (i.e.\ also
$|\delta|$) are small compared to $\hbar\omega$. We arrive at the approximate effective 
Hamiltonian (\ref{eq:Heff}), but with $v_\ell$ given by Eq.~(\ref{eq:well}) and with the 
effective tunneling matrix element reading \cite{Zak93,EckardtHolthaus07} 
\be\label{eq:JeffNu}
J_\text{eff}=J\Jb_\nu\bigg(\frac{K}{\hbar\omega}\bigg).
\ee
For small arguments the Bessel function behaves like
$\Jb_\nu(x)\simeq\frac{1}{|\nu|!} \big[\mathrm{sgn}(\nu) \frac{x}{2}\big]^{|\nu|}$, 
so that for $\nu\ne0$ the effective tunneling matrix element vanishes for $K/\hbar\omega=0$. 
This reflects the fact that for a strong potential tilt $\nu\hbar\omega\gg J$ tunneling is 
suppressed. However, switching on a finite driving strength $K/\hbar\omega$ the effective 
tunneling matrix element acquires finite values such that coherent ``photon''-assisted 
tunneling is induced by the periodic force.

The fact that the localized Wannier-Stark eigenstates of the tight-binding model described by
$\Ho_\text{eff}$ with a finite tilt $\delta$ are known explicitly in the absence of trapping 
potentials and interactions, allows for an analytical description of the dynamics of an 
initially localized wave packet in the shaken tilted lattice (\onlinecite{ThommenEtAl02,
ThommenEtAl04a,ThommenEtAl04b}, see also \onlinecite{KolovskyKorsch10,KudoMonteiro11,
CreffieldSols11}), as it has been investigated in different experiments to be described in the 
following paragraphs.

``Photon''-assisted tunneling described by the effective tunneling matrix element
(\ref{eq:JeffNu}) with $\nu=1$ and $\nu=2$ has been observed in Arimondo's group in Pisa
\cite{SiasEtAl08} from the coherent expansion of a Bose condensate of Rb-87 atoms in a
one-dimensional lattice, where both the static and the sinusoidal force where created by 
lattice acceleration [a seeming discrepancy between the measured data with the prediction
(\ref{eq:JeffNu}) was later resolved by taking into account the initial extent of the 
condensate when extracting the effective tunneling matrix element \cite{CreffieldEtAl10}]. 

At the same time a similar experiment has been conducted in Tino's group in Florence with 
Sr-88 atoms in an optical cosine lattice, where the static force was given by gravitation and 
the periodic force was realized via lattice shaking \cite{IvanovEtAl08}. The experimentalists 
observed ballistic spreading for resonant forcing with $\hbar\omega=n\Delta$ and $n=1$, 2, 3 
and 4. The authors attribute the resonances to tunneling processes between lattice sites at 
distance $nd$. Since for the used lattice depth of $V_0/E_R=20$ the matrix elements for
next-nearest-neighbor tunneling are negligible, this interpretation suggests that atoms were 
loaded also into excited Bloch bands of the lattice, where also longer-ranged tunneling matrix 
elements matter. An alternative mechanism leading to such resonances would be $n$th-order 
tunneling processes, where $n$ particles tunnel to a neighboring site via virtual intermediate  
non-resonant states. 

In a later experiment by the same group the atoms were loaded into the lowest band of a tilted 
lattice and the impact of the small effective lattice tilt $\delta=\Delta-\hbar\omega$ was 
explored, as it appears in the effective Hamiltonian (\ref{eq:Heff}) through $v_\ell$
[Eq.~(\ref{eq:well})]. The experimentalists observed a large-amplitude breathing dynamics of 
the initially localized atom cloud at the small effective Bloch frequency $\delta/(2\pi\hbar)$ 
\cite{AlbertiEtAl09}. For the lowest effective Bloch frequency of approximately $0.26$ Hertz a 
breathing amplitude of about one millimeter was observed. The driven system retained coherence 
over \emph{macroscopic} times and distances. Since the atom cloud was hot, with the momentum 
distribution smeared out over the whole first Brillouin zone, no center-of-mass Bloch 
oscillations were observed. However, the thermal nature of the initial state did not destroy 
the coherent breathing dynamics. 

In an experiment by N\"agerl's group in Innsbruck also center-of mass oscillations at the 
effective Bloch frequency $\delta/(2\pi\hbar)$ were observed \cite{HallerEtAl10}. In this 
experiment with bosonic Cs-133 atoms in a one-dimensional optical lattice the static tilt was 
given by gravitation and the sinusoidal force was realized using an oscillating magnetic-field 
gradient. Since the atom cloud possessed a peaked quasimomentum distribution, it acquired also 
a large-amplitude center of mass oscillation (\emph{super Bloch oscillations},
\onlinecite{KolovskyKorsch10}), described by the group velocity of the effective Hamiltonian 
at the oscillating quasimomentum peak. The amplitude of the oscillations was found to be 
determined by $J_\text{eff}$ as given by Eq.~(\ref{eq:JeffNu}) with $\nu=1$. Their phase, as 
it is determined by the quasimomentum shift acquired while the forcing is switched on, was 
controlled by the time when during the driving period the forcing was switched on abruptly. 

Another possibility to resonantly induce effective coherent tunneling in a strongly tilted 
lattice is a sinusoidal modulation of the lattice depth, which is captured by a modulation of 
the tunneling matrix element in the tight-binding Hamiltonian~(\ref{eq:DBH}), 
\be\label{eq:Jamp}
J\to J(t) = \sum_{\mu=-\infty}^\infty J^{(\mu)} e^{i\mu\omega t} \simeq J 
			+  \Delta J\cos(\omega t), 
\ee
with $J^{(-\mu)}=J^{(\mu)*}$. Additionally, there is also a weak periodic modulation of the 
interaction parameter, $U\to U(t)\simeq U - \Delta U \cos(\omega t)$. The Wannier-Stark tilt is 
captured by 
\be\label{eq:vw}
w_\ell= \frac{x_\ell}{d}\nu\hbar\omega,\qquad
v_\ell = v^\text{tr}_\ell + \frac{x_\ell}{d}\delta.
\ee
Integrating out the strong potential tilt included in $w_\ell$ by a gauge transformation
[Eqs.~(\ref{eq:Uchi}) and (\ref{eq:chi})], one arrives at a Hamiltonian $\Ho'(t)$ of the
form~(\ref{eq:Hprime}), with $J$ and $U$ replaced by $J(t)$ and $U(t)$ and Peierls
phases~(\ref{eq:Peierls}) determined by the quasimomentum shift
\be
\ba(t)=[-\frac{1}{d}\nu\omega t + a_0]{\bm e}_x.
\ee
Once again, one can approximate $\Ho'(t)$ by its time average, giving the effective Hamiltonian
(\ref{eq:Heff}), where $v_\ell$ is given by Eq.~(\ref{eq:vw}) and where for the choice of gauge $a_0=0$ the effective tunneling matrix element reads
\be\label{eq:Jeffamp}
J_\text{eff} = J^{(-\nu_{\ell'\ell})} 
\simeq  \delta_{\nu,0}J+ \frac{\delta_{\nu,1}+\delta_{\nu,-1}}{2}\,\Delta J,
\ee
with integer $\nu_{\ell'\ell}\equiv (w_{\ell'}-w_\ell)/\hbar\omega$ for tunneling from $\ell$ 
to $\ell'$. Thus, as long as higher harmonics are negligible in the modulated tunneling matrix 
element $J(t)$, the forcing allows for single-``photon'' processes bridging energy differences 
of $\pm\hbar\omega$.

Such ``photon''-assisted tunneling via a modulation of the lattice depth has been observed in 
Florence \cite{AlbertiEtAl10}, where \emph{inter alia} the selectivity to single-``photon'' 
processes has also been used in order to selectively induce and observe tunneling processes 
between next-nearest and next-next nearest lattice sites. Subsequently, the scheme was employed 
for a precision measurement of the gravitational acceleration \cite{PoliEtAl11}. 

This effect has been combined with strong interactions both with spinless bosons in a tilted 
lattice by Greiner's group at Harvard \cite{MaEtAl11} and with spin-1/2 bosons in a tilted 
double well by Bloch's group in Munich \cite{ChenEtAl11}. In these experiments the on-site 
interaction $U$ is large compared to the tunneling parameter $J$ and comparable to the lattice 
tilt $\Delta$, so that the resonance condition (\ref{eq:Delta}) has to be modified to include 
also the change of interaction energy associated with a tunneling process. In this way the 
effective tunneling matrix element becomes occupation dependent. In Greiner's lab, this effect 
was employed to measure the Hubbard energy $U$ as well as occupation-number-dependent 
corrections to it, as they arise from perturbative admixtures of excited Bloch bands. Moreover, 
it has been used to control (a finite-size precursor of) a phase transition in the effective 
spin model that the same group had realized already in an earlier experiment with spinless 
bosons in a tilted lattice \cite{SimonEtAl11}. In the Bloch experiment also second-order 
tunneling via non-resonant intermediate states was observed at the resonance condition
$\hbar\omega=2\Delta$ (see also \onlinecite{EckardtEtAl05a}) and the modulation was used to 
control superexchange processes (see also \onlinecite{MentinkEtAl14, BukovEtAl16}). 

We note in passing that ``photon''-assisted tunneling has recently also been observed in a 
lattice of optical wave guides \cite{MukherjeeEtAl15}. Experiments that use
``photon''-assisted tunneling induced by a moving secondary lattice for the purpose of 
engineering artificial magnetic fields will be reviewed in Sec.~\ref{sec:msl} below.

\subsection{Dynamic control of the bosonic superfluid-to-Mott-insulator transition}

\begin{figure}[t]
\includegraphics[width=1\linewidth]{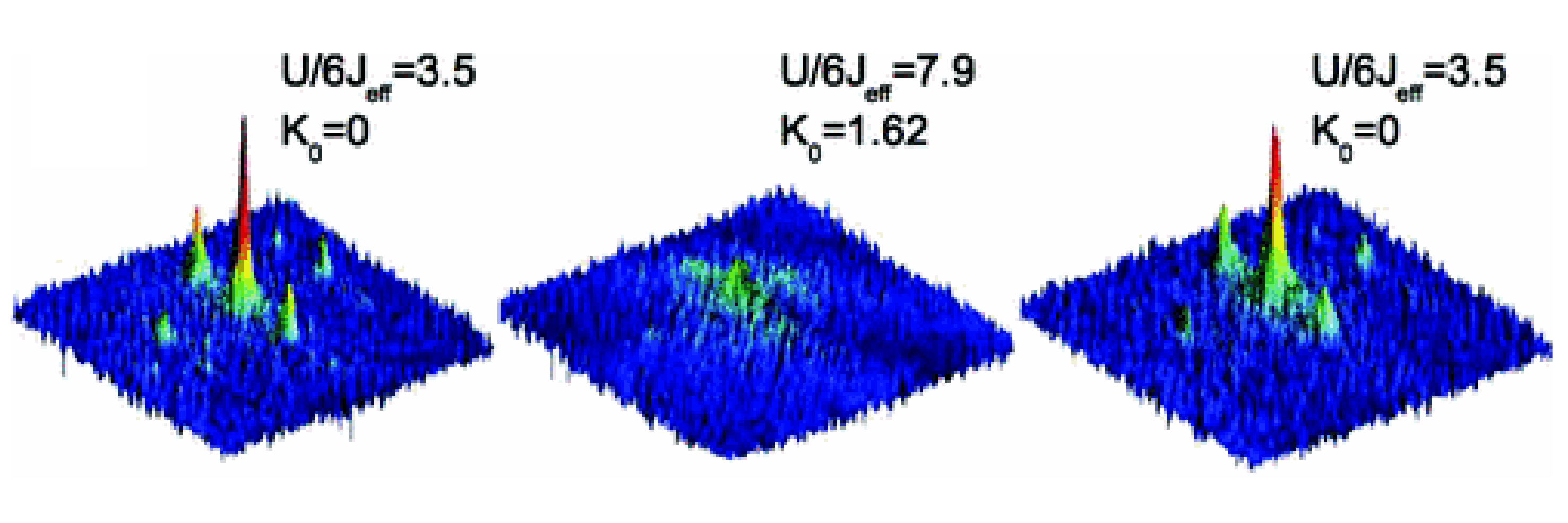}
\centering
\caption{\label{fig:SFMI} Dynamically induced Superfluid-to-Mott-insulator transition in a 
shaken cubic optical lattice. Two-dimensional projection of the momentum distribution obtained 
from time-of-flight absorption imaging at three different times during the experimental 
protocol: before ramping up the driving strength $K_0=K/\hbar\omega$, after $K_0$ has been 
ramped up linearly (middle), after $K_0$ has been ramped down again (right). The loss and
re-appearance of sharp peaks indicates that the system approximately followed a many-body 
Floquet state undergoing a quantum phase transition from a superfluid to a Mott insulator and 
back. (taken from \onlinecite{ZenesiniEtAl09})} 
\end{figure}

When deriving the approximate effective Hamiltonian (\ref{eq:Heff}), in the previous sections 
it was assumed that the driving frequency is large compared to tunneling and interaction 
parameters. But it was not required that the interactions are weak compared to the kinetic 
energy. Therefore, it is possible to control a lattice system also in the strong coupling 
regime by means of periodic forcing. This has been exploited in an experiment in Pisa
\cite{ZenesiniEtAl09}, where the transition between a bosonic superfluid and a Mott-insulator 
state has been induced by means of lattice shaking. This experiment followed a proposal by 
\onlinecite{EckardtEtAl05b} (see also \onlinecite{EckardtHolthaus07}).

The bosonic Hubbard model possesses two different ground-state phases, a gapless compressible 
superfluid phase, with the particles being delocalized, and a gapped incompressible
Mott-insulator phase, where an integer number of particles is localized at every lattice site 
by strong repulsive interactions \cite{FisherEtAl89}. In a trapped optical lattice system 
extended Mott-insulator regions form when the ratio between Hubbard interaction and tunneling 
parameter, $U/J$, exceeds a critical value \cite{JakschEtAl98}, which for the three-dimensional 
cubic lattice at unit filling is given by $(U/J)_c\approx 29.3$ 
\cite{CapogrossoSansoneEtAl07,TeichmannEtAl09}. This transition has been observed the first 
time in a seminal experiment with spinless bosons in a cubic optical lattice by
\onlinecite{GreinerEtAl02} in Munich, where the ratio $U/J$ was increased by ramping up the 
lattice depth $V_0$. As a signature of the transition the experimentalists observed the 
disappearance of sharp peaks in the momentum distribution, as they characterize the superfluid 
phase, as well as their reappearance, when the lattice depth was ramped down again. 

In the Pisa experiment lattice shaking was employed to lower the effective tunneling parameter
(\ref{eq:Jeff}) with respect to the interaction strength $U$, which is not altered by the 
lattice shaking. In order to modify the tunneling matrix element in all three directions of a 
cubic lattice, the forcing was applied along the diagonal direction. When the shaking amplitude 
was ramped up smoothly, the sharp momentum peaks characterizing the superfluid ground state 
disappeared once $U/J_\text{eff}$ became sufficiently large. The peaks reappeared, when the 
forcing was ramped down again (see Fig.~\ref{fig:SFMI}). The interpretation of this experiment 
is that the system followed approximately a many-body Floquet state that, in response to the 
variation of the driving amplitude, underwent a transition from a superfluid to a Mott 
insulator and back. This experiment demonstrates on the one hand that time-periodic forcing is 
a suitable tool also for the {manipulation of strongly interacting many-body systems} and their 
interaction-driven physics. On the other hand, it also is an example of \emph{adiabatic state 
preparation} in a time-periodically driven system. Such ``adiabatic'' processes in driven
many-body systems will be discussed in more detail in section \ref{sec:FloquetPicture} below, 
where we will point out that they actually correspond to a complex mixture of adiabatic and 
diabatic processes in an extended Hilbert space.

\subsection{\label{sec:frustration}Kinetic frustration}

Controlling the spreading of a Bose condensate, Bloch oscillations, or even the
superfluid-to-Mott-insulator transition, all these experimentally observed effects discussed 
above clearly show that periodic forcing is a suitable tool for controlling many-body systems 
of ultracold atoms in optical lattices. These effects have, however, been achieved also without 
periodic forcing, e.g.\ by varying the depth of the optical lattice, leading to an exponential 
suppression of the tunneling matrix element, or by tuning a Wannier-Stark tilt. But periodic 
forcing can also be used to engineer systems with qualitatively new properties. One possibility 
is to effectively modify not only the amplitude, but also the sign or, more generally, the 
phase of tunneling matrix elements.  

In the one-dimensional driven tight-binding chain, with effective dispersion relation
$\varepsilon_\text{eff}(k_x)=-2J_\text{eff}\cos(dk_x)$, a sign change of the effective 
tunneling matrix element $J_\text{eff}$ does not lead to qualitatively new physics. The 
resulting inversion of the dispersion relation can be compensated by a shift in quasimomentum 
by $\Delta k=\pi/d$, which corresponds to a gauge transformation and leaves also the 
interactions unchanged. This argument generalizes to other bipartite lattice geometries (like 
square, hexagonal, or cubic), where a sign change of the tunneling matrix element can be 
compensated by redefining the sign of the Wannier orbital on every other lattice site. However, 
for a non-bipartite lattice (like the triangular or the Kagom\'e lattice), the inversion of the 
tunneling matrix element does not simply correspond to a gauge transformation, but leads to a 
geometrically frustrated tunneling kinetics: 
A negative tunneling parameter $J_\text{eff}<0$, corresponding to a positive tunneling matrix 
element $-J_\text{eff}>0$, favors the wave function to change sign from one lattice site to the 
other. Thus, given, e.g., three sites arranged in a triangular plaquette, it is not possible 
anymore to minimize the kinetic energy at each of the three tunneling bonds at the same time. 
Especially in combination with strong interactions, such kinetic frustration can give rise to 
intriguing behavior. 

\begin{figure}[t]
\includegraphics[width=1\linewidth]{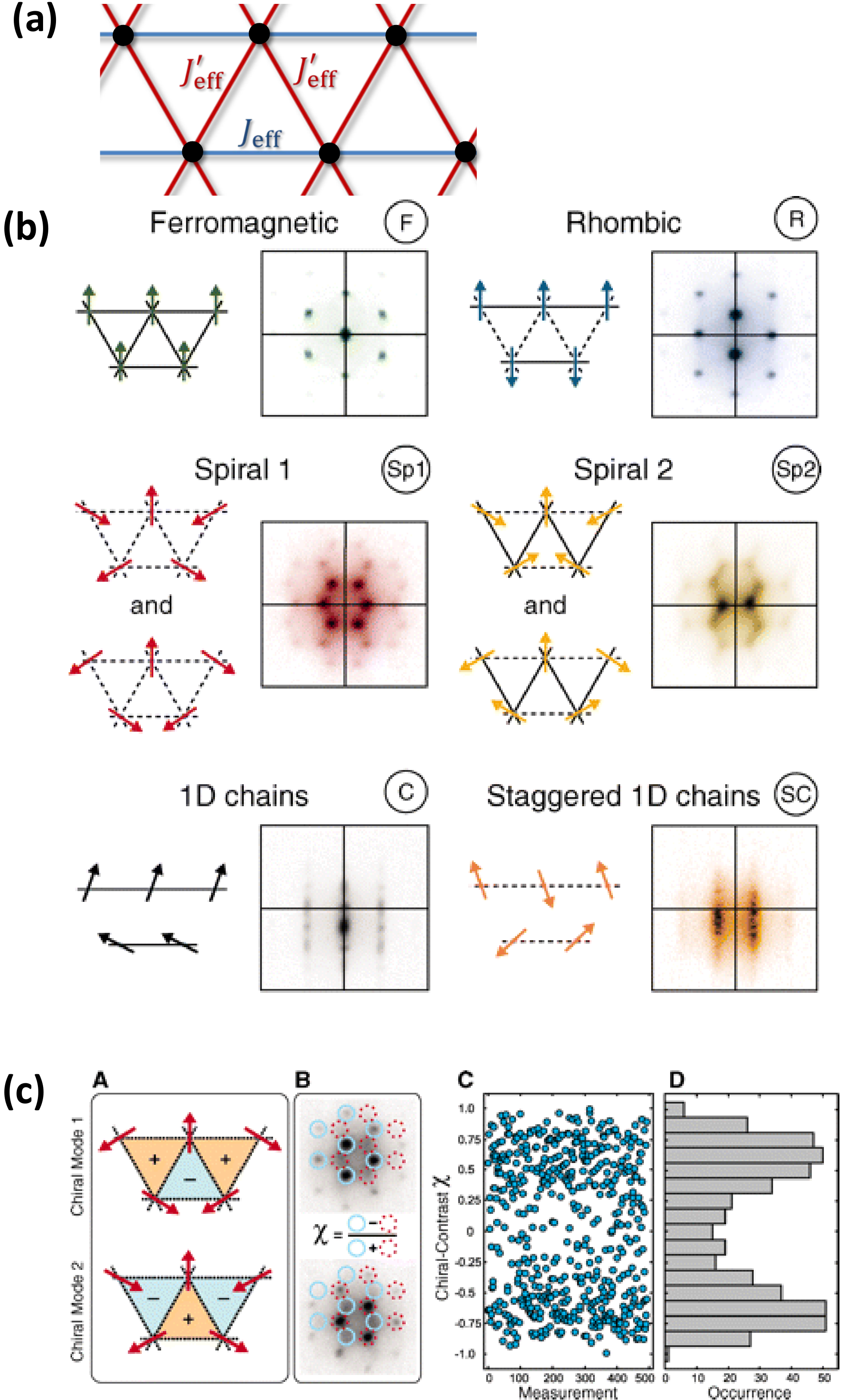}
\centering
\caption{\label{fig:tri} (a) Anisotropic triangular lattice with effective tunneling 
parameters $J_\text{eff}$ and $J'_\text{eff}$. (b) Momentum distribution averaged over many 
measurements (shots) and corresponding pattern of the condensate phase (indicated by direction 
of arrow) for different $J_\text{eff}$ and $J_\text{eff}'$ (dashed/solid lines indicate
positive/negative tunneling parameters). (c) Spontaneous time-reversal symmetry breaking for
$J_\text{eff}=J'_\text{eff}<0$. Each of the two spatial configurations of the condensate phase 
shown in A breaks time-reversal symmetry. In absorption images the two are distinguished by the 
position of the measured peaks, indicated by dotted or solid circles in B. The contrast 
between both configurations $\chi$ varies from shot to shot (C) giving a bimodal distribution
(D), so that typically only one of the two configurations appears spontaneously.  
[(b) and (c) taken from \onlinecite{StruckEtAl11}]} 
\end{figure}

A sign inversion of the effective tunneling parameter can be achieved via periodic forcing 
\cite{EckardtEtAl10}. A two-dimensional lattice that is shaken along a circular orbit 
experiences the inertial force 
\be\label{eq:Fcirc}
\bF(t)=F_\omega[\cos(\omega t){\bm e}_x+\sin(\omega t){\bm e}_y].
\ee 
In the high-frequency regime its dynamics can be described by the approximate effective 
Hamiltonian (\ref{eq:Heff}) with the isotropic tunneling parameter given by Eq.~(\ref{eq:Jeff}),
where $K=dF_\omega$ with lattice constant $d$. The effective tunneling parameter becomes 
negative at $K/\hbar\omega\approx 2.4$ and assumes a minimal value of
$J_\text{eff}\approx -0.4 J$ at $K/\hbar\omega\approx3.8$. On the single-particle level, after 
an inversion of the tunneling matrix elements, the dispersion relation of the triangular 
lattice $\varepsilon_\text{eff}(\bk)$ possesses two inequivalent minima $\bk=\pm\bq$. For a 
Bloch wave function 
$\psi_\ell=M^{\!-1/2}\exp(i\varphi_\ell)$ with $\varphi_\ell=\bk\cdot\br_\ell$ on the lattice, 
where $M$ denontes the total number of lattice sites, the effective kinetic energy is given by 
\be
\varepsilon_\text{eff}(\bk)
	=-\frac{J_\text{eff}}{M}\sum_{\la\ell'\ell\ra}\cos(\varphi_{\ell'}-\varphi_\ell) 
	=-J_\text{eff}\sum_{\bm d}\cos({\bm d}\cdot\bk),
\ee
where the second sum runs over the six vectors $\bm d$ that connect each lattice site with its 
nearest neighbors. The phases $\varphi_\ell$ play the role of coupled classical rotors. For 
antiferromagnetic coupling, $J_\text{eff}<0$, this energy becomes minimal for the two spiral 
phase patterns denoted ``Spiral 1'' in Fig.~\ref{fig:tri}(b), characterized by $\bk=\pm \bq $ 
with $\bq=(q_x,0,0)$ and $q_x=4\pi/(3d)$. 

Considering many, weakly repulsive, spinless bosons, the ground state of the effective 
Hamiltonian corresponds to a Bose condensate in one of the two minima of the effective 
dispersion relation [states involving both quasimomenta $\bq$ and $-\bq$, either in a coherent 
supersposition or by forming a fragmented condensate, are disfavored by repulsive 
interactions \cite{EckardtEtAl10}]. This form of spontaneous time-reversal symmetry breaking 
has been observed in a system of weakly interacting spinless bosons in a triangular lattice of 
one-dimensional tubes in Sengstock's group in Hamburg \cite{StruckEtAl11}, see
Fig.~\ref{fig:tri}(c).

Extending the scheme to elliptical forcing causes an anisotropic modification of tunneling, 
since the amplitude of the forcing $K$ acquires a directional dependence. In a triangular 
lattice this allows for creating the pattern of tunneling matrix elements depicted in
Fig.~\ref{fig:tri}(a). The parameter space spanned by $J_\text{eff}$ and $J_\text{eff}'$ has 
been explored in the Hamburg experiment. Fig.~\ref{fig:tri}(b) shows the measured momentum 
distributions, which feature peaks at the expected positions corresponding to the sketched 
phase patterns. The degree of kinetic frustration is basically controlled by
$J_\text{eff}/|J'_\text{eff}|$. For $J_\text{eff}=0$, the remaining $J'_\text{eff}$ bonds form 
a bipartite rhombic lattice, which does not feature frustration and favors a staggered
N\'eel-type order of the phases $\varphi_\ell$  [denoted "rhombic" in Fig.~\ref{fig:tri}(b)] 
for antiferromagnetic coupling $J'_\text{eff}<0$. Switching on a finite antiferromagnetic
$J_\text{eff}=-\gamma|J'_\text{eff}|<0$ leads to frustration for either sign of $J'_\text{eff}$,
and causes spontaneous time-reversal symmetry breaking when $\gamma>\gamma_c$. The 
corresponding phase patterns are denoted by ``Spiral 1'' and ``Spiral 2'' for $J'_\text{eff}<0$ 
and $J'_\text{eff}>0$, respectively. The critical parameter is roughly given by
$\gamma_c\approx 0.5$, where the single-particle dispersion relation develops two minima, 
though interaction-induced quantum fluctuations are expected to shift it to slightly larger 
values \cite{EckardtEtAl10}. 

The idea of achieving kinetic frustration via lattice shaking is of interest mainly for 
bosons. For fermions, an inversion of tunneling matrix elements results already from a
particle-hole transformation, 
so that kinetic frustration will naturally appear when the Fermi energy becomes sufficiently 
large. Apart from the triangular lattice, the scheme can be used to induce kinetic frustration 
also in other non-bipartite lattice geometries. For an optical Kagom\'e lattice, as it has been 
realized experimentally recently \cite{JoEtAl12}, the impact of kinetic frustration would be 
even more drastic. After inverting the sign of the tunneling parameters the lowest of the three 
bands will be completely flat, so that even weak interactions will have a major impact on the 
ground state \cite{HuberAltman10}. Also one-dimensional chains, like the saw-tooth
\cite{HuberAltman10} or zig-zag \cite{GreschnerEtAl13} lattice, can acquire kinetic frustration 
in response to lattice shaking as well as non-bipartite three-dimensional lattice geometries, 
such as pyrochlore. Kinetic frustration enhances the role of interactions not only in the 
extreme case of lattice geometries acquiring a flat lowest band. In the triangular lattice the 
critical interaction strength for the formation of a Mott-insulator will be reduced
\cite{EckardtEtAl10} and it can even become zero in the zig-zag chain \cite{GreschnerEtAl13}. 
Note that the system can form a chiral Mott insulator, with spontaneously broken time-reversal 
symmetry breaking appearing in the particle-hole fluctuations
\cite{GreschnerEtAl13,ZaletelEtAl14}. 

It is also an interesting perspective to explore the interplay of kinetic frustration with very
strong interactions. In the limit of hard-core bosons the effective Hamiltonian (\ref{eq:Heff}) 
can be mapped to a quantum spin-1/2 $XY$ model \cite{EckardtEtAl10}
\be\label{eq:Hspin}
\Ho_\text{eff}=-J_\text{eff}\sum_{\la\ell'\ell\ra} \hat{S}_{\ell'}^{+}\hat{S}_{\ell}^{-}
		= -J_\text{eff}\sum_{\la\ell'\ell\ra}
	\Big(\hat{S}_{\ell'}^{x}\hat{S}_{\ell}^{x}+\hat{S}_{\ell'}^{y}\hat{S}_{\ell}^{y}\Big).
\ee
Here the $\hat{S}_\ell$ denote standard spin operators acting on the pseudo-spin degree of 
freedom spanned by the two states ``there is a boson'' ($\uparrow$) and ``there is no boson''
($\downarrow$). In the experiment mentioned above, reaching this regime would requires a 
further confinement perpendicular to the lattice, making the system effectively 
two-dimensional. For non-bipartite lattices and $J_\text{eff}<0$ the Hamiltonian
(\ref{eq:Hspin}) describes frustrated quantum antiferromagnetism. The ground-state
(and low-temperature) regime of such frustrated quantum magnets can give rise to intriguing 
physics, like the formation of topological or critical spin liquids. However, the theoretical 
prediction of the nature of the ground state is typically a hard problem 
\cite{MoessnerRamirez06, Sachdev08,Balents10}. Possibly, future experiments simulating the 
Hamiltonian (\ref{eq:Hspin}) in shaken optical lattices of various geometries could provide 
useful information concerning this issue. 
Here a promising feature is that the model (\ref{eq:Hspin}) is based on easy-to-cool motional 
bosonic degrees of freedom, with the coupling on the order of the tunneling matrix element
$J_\text{eff}\sim J$. Without a lattice, bosonic systems have been cooled down to entropies per 
particle as low as $0.001k_\text{B}$ \cite{OlfEtAl15}. This contrasts with optical-lattice 
spin systems based on a Mott insulator of spin-1/2 fermions, with small superexchange coupling 
$\sim J^2/U\ll J$ between neighboring spins. For spin-1/2 fermions entropies per particle of 
about $0.6k_\text{B}$ in the Mott-insulating state \cite{GreifEtAl13, HartEtAl15, BollEtAl16} 
and $0.04k_\text{B}$ in a system without lattice \cite{KuEtAl12} have been achieved. Moreover, 
close analogies between ground and low-energy states of frustrated XY and Heisenberg 
antiferromangets might permit to shed light also on the physics of the latter 
\cite{LaeuchliMoessner15}.

\subsection{\label{sec:HFgauge}Artificial magnetic fields -- High-frequency schemes}
Inverting the sign of the tunneling matrix elements can be viewed as a special case of a more 
general scheme where the effective matrix element for tunneling from $\ell$ to $\ell'$ 
acquires a \emph{phase},
\be
J^\text{eff}_{\ell'\ell}=|J^\text{eff}_{\ell'\ell}|\,e^{i\theta^\text{eff}_{\ell'\ell}}.
\ee
Such effective Peierls phases $\theta^\text{eff}_{\ell'\ell}$ play the role of a vector 
potential. The tight-binding representation of a vector potential ${\bm A}(\br)$ is, according 
to the Peierls substitution, given by
$\theta_{\ell'\ell}=\frac{1}{\hbar}\int_{\br_{\ell}}^{\br_{\ell'}}\!\rd\br\cdot{\bm A}(\br)$, 
where we have absorbed the charge in the definition of ${\bm A}$ so that it carries the 
dimension of a momentum and where the integration is taken along a straight line. In the last 
years, such effective Peierls phases have been realized by means of periodic forcing in several 
experiments \cite{AidelsburgerEtAl11,StruckEtAl12,StruckEtAl13,AidelsburgerEtAl13,MiyakeEtAl13,
AtalaEtAl14,AidelsburgerEtAl15,KennedyEtAl15}.

\begin{figure}[t]
\includegraphics[width=1\linewidth]{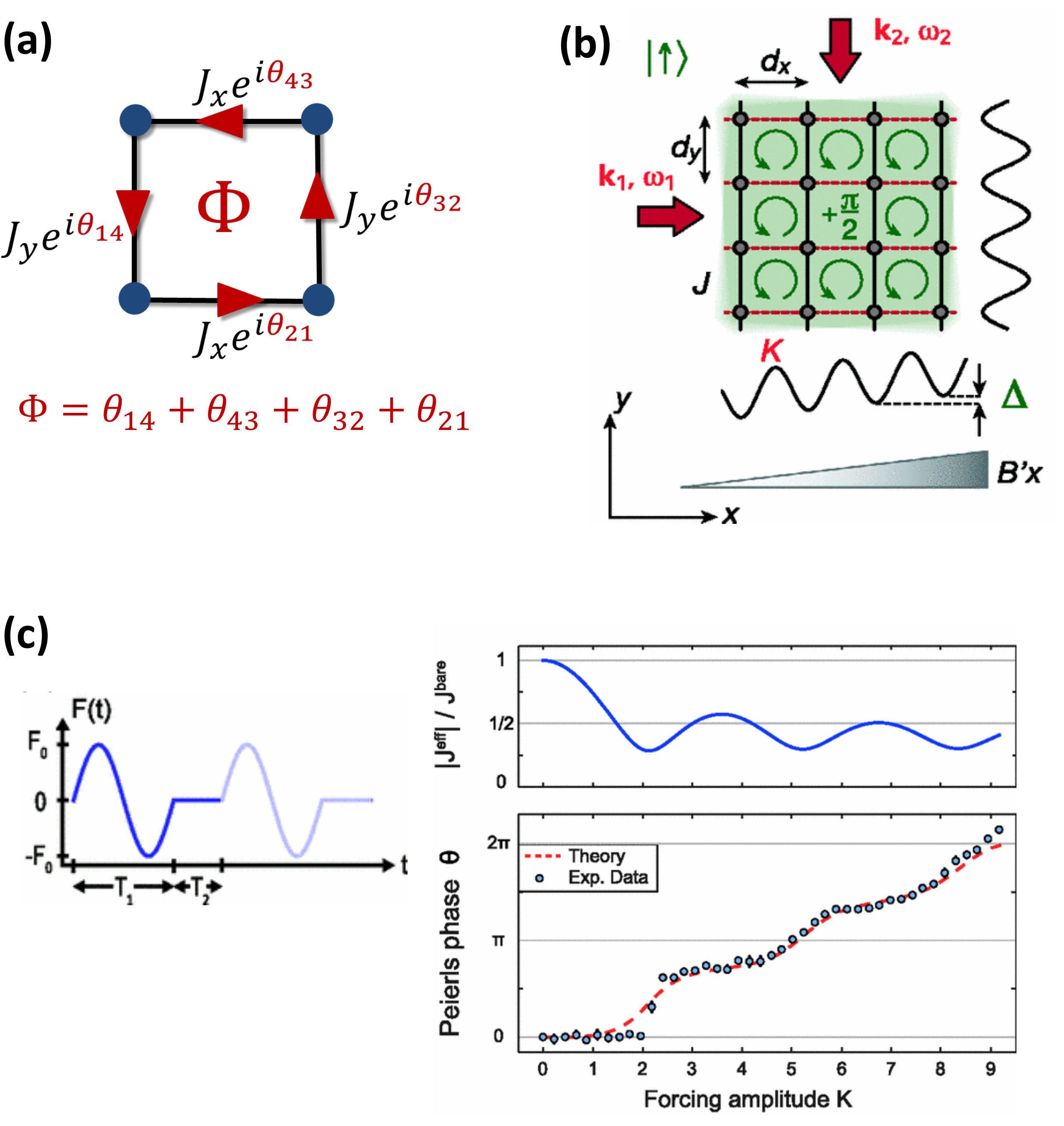}
\centering
\caption{\label{fig:flux} (a) The dimensionless magnetic flux $\Phi$ piercing a lattice 
plaquette equals the sum of the Peierls phases $\theta_{\ell'\ell}$ picked up when moving 
around it in positive direction once. (b) Moving-secondary-lattice scheme for creating a 
a homogeneous flux configuration. (d) Assymmetric lattice shaking (left) gives rise to complex 
effective tunneling matrix elements (right), plotted versus $K=F_0dT_1/T$ (in units of
$\hbar\omega$) for $T_1/T_2=2.1$. [taken from: (b) \onlinecite{AidelsburgerEtAl11}, (c) from
\onlinecite{AidelsburgerEtAl13}, (d) \onlinecite{StruckEtAl12}]} 
\end{figure}

Of particular interest is the situation where the Peierls phases describe a finite effective 
magnetic flux $\Phi^\text{eff}_P$ through a lattice plaquette $P$. It is defined as the 
dimensionless Aharonov-Bohm-like phase $\Phi^\text{eff}_P= \sum_P \theta^\text{eff}_{\ell'\ell}$
obtained by summing over the Peierls phases picked up when tunneling once around the plaquette 
in positive direction, as depicted in Fig.~\ref{fig:flux}. The plaquette flux is defined modulo 
the dimensionless magnetic flux quantum of $2\pi$ only, it is gauge invariant and plays the 
role of the magnetic field (flux density) in continuous systems. The creation of effective 
plaquette fluxes by means of periodic forcing turned out to be a powerful method for the 
creation of \emph{artificial} (or \emph{synthetic}) magnetic fields for charge-neutral 
particles in optical lattices (other schemes rely on laser-dressing of internal atomic degrees
of freedom, \onlinecite{DalibardEtAl11,GoldmanEtAl14}). In this way extremely strong fields of 
the order of the maximum possible flux of $\pi$ can be achieved. To put this in perspective, 
for an electron a flux of $\pi$ through the hexagonal plaquette of graphene with area
$A_\text{hex}\approx5.2\text{\AA}^{\!2}$ would correspond to the enormous magnetic field 
strength $B=\pi \hbar/(eA_\text{hex}) \approx 3.9\cdot10^4 \text{T}$, which is more than two 
orders of magnitude larger than the real magnetic fields that can be achieved in the laboratory. 

The experiments to be discussed in the following are based on tailoring on-site potentials of 
the form 
\be
w_\ell(t) = w^\text{dr}_\ell(t) + \nu_\ell\hbar\omega,
\ee
appearing in the Hamiltonian (\ref{eq:DBH}), with time-periodic potential modulation
$w^\text{dr}_\ell(t)=w^\text{dr}_\ell(t+T)$ of zero average,
$\frac{1}{T}\int_0^T\!\rd t w^\text{dr}_\ell(t)=0$, and possibly also a static part with 
integers $\nu_{\ell}$. In the high-frequency regime, a system that is driven like that will 
again be described by an effective Hamiltonian of the form (\ref{eq:Heff}), with the tunneling 
matrix elements $J^\text{eff}_{\ell'\ell}$ given by Eq.~(\ref{eq:Jeff_ell}) depending on
$w_\ell(t)$ as specified by Eq.~(\ref{eq:chi}). 

Before reviewing specific schemes and experiments, let us identify necessary conditions for the 
creation of artificial gauge fields (\onlinecite{HaukeEtAl12b}, a general discussion of 
symmetries of the effective Hamiltonian is furthermore given by \onlinecite{KitagawaEtAl10}). 
For that purpose, we choose the gauge constant $\chi_{0\ell}$ in Eq.~(\ref{eq:chi}) such that
$\chi_\ell(t)=\chi^\text{dr}_\ell(t)-\nu_\ell\omega t+\gamma_\ell$, with
$\chi^\text{dr}_\ell(t)$ having zero average,
$\frac{1}{T}\int_0^T\!\rd t \chi^\text{dr}_\ell(t)=0$, and with yet undetermined constants
$\gamma_\ell$ representing the gauge freedom. The imaginary part of the 
effective tunneling matrix element $J^\text{eff}_{\ell'\ell}$ is given by
$I_{\ell'\ell}= -\frac{J}{T}\int_0^T\!\rd t\,
\sin\big(\chi^\text{dr}_{\ell'\ell}(t)-\nu_{\ell'\ell}\omega t 
+ \gamma_{\ell}-\gamma_{\ell'}\big)$, where
$\chi_{\ell'\ell}^\text{dr}=\chi_{\ell'}^\text{dr}-\chi_{\ell}^\text{dr}$ 
and $\nu_{\ell'\ell}=\nu_\ell'-\nu_\ell$. If one can find gauge constants $\gamma_{\ell}$ such 
that all $I_{\ell'\ell}$ vanish, one cannot create plaquette fluxes
$\Phi^\text{eff}_P\ne 0,\pi$ that break time reversal symmetry (where the special case
$\Phi_P^\text{eff}=\pi$ corresponds to the situation of kinetic frustration discussed above). 

Let us first discuss the case, where the static potential off-sets vanish, $\nu_{\ell'\ell}=0$. 
One can identify two temporal symmetries of the relative potential modulations
$w^\text{dr}_{\ell'\ell}\equiv w^\text{dr}_{\ell'}-w^\text{dr}_{\ell}$ that imply
$I_{\ell'\ell}=0$ for the choice $\gamma_\ell=\gamma_\ell'=0$. These are the
\emph{local reflection symmetry} 
\be\label{eq:lr}
w^\text{dr}_{\ell'\ell}(t-\tau_{\ell'\ell})=w^\text{dr}_{\ell'\ell}(-t-\tau_{\ell'\ell})
\qquad \forall \la\ell'\ell\ra
\ee 
with respect to times $\tau_{\ell'\ell}$ defined individually on each local bond $(\ell'\ell)$, 
and the \emph{shift symmetry} 
\be\label{eq:shift}
w^\text{dr}_{\ell'\ell}(t)=-w^\text{dr}_{\ell'\ell}(t-T/2) \qquad \forall \la\ell'\ell\ra.
\ee 
Either of these symmetries implies that the effective Hamiltonian preserves time-reversal 
symmetry. Note that precisely these symmetries are also known to prevent ratchet-type transport
\cite{FlachEtAl00,DenisovEtAl07}.\footnote{Such directed transport has also been studied 
experimentally with atomic quantum gases in driven optical lattices in Weitz's group in Bonn 
\cite{SalgerEtAl09,SalgerEtAl13}.} A sinusoidal potential modulation obeys both symmetries.

If additional to the potential modulations also finite potential off-sets $\nu_{\ell'\ell}\ne0$ 
are created, so that the driving has to induce ``photon'' assisted tunneling, the above 
symmetries are not enough to enforce time-reversal symmetry $I_{\ell'\ell}=0$. Instead this can 
be achieved by choosing $\gamma_\ell=-\nu_\ell\omega \tau$, if the \emph{global reflection 
symmetry} 
\be\label{eq:gr}
w^\text{dr}_{\ell'\ell}(t-\tau)=w^\text{dr}_{\ell'\ell}(-t-\tau)
\qquad \forall \la\ell'\ell\ra
\ee
is fulfilled with respect to a globally defined time $\tau$. 

Involving ``photon'' assisted tunneling against non-zero potential off-sets
$\nu_{\ell'\ell}\hbar\omega$ poses less constraints for the creation of artificial gauge 
fields, since only the global reflection symmetry (\ref{eq:gr}) has to be broken. As a 
consequence, already sinusoidal forcing
\be
w^\text{dr}_\ell(t)=K\sin(\omega t-\varphi_\ell)
\ee
can produce plaquette fluxes $\Phi^\text{eff}_P\ne 0,\pi$, provided the driving phase
$\varphi_\ell$ varies from site to site. This has been proposed theoretically
\cite{Kolovsky11,BermudezEtAl11} and demonstrated in a series of beautiful 
experiments in the groups of Bloch in Munich \cite{AidelsburgerEtAl11,AidelsburgerEtAl13,
AtalaEtAl14, AidelsburgerEtAl15} and Ketterle at MIT \cite{MiyakeEtAl13,KennedyEtAl15}. 
For $\gamma_\ell=0$, one finds 
\be\label{eq:JeffPA}
J^\text{eff}_{\ell'\ell} = J\Jb_{\nu_{\ell'\ell}}\Big(\frac{K_{\ell'\ell}}{\hbar\omega}\Big)
			\,e^{i\nu_{\ell'\ell}\varphi_{\ell'\ell}},
\ee
with amplitude $K_{\ell'\ell}=2K\sin(\varphi_{\ell'}-\varphi_\ell)$ and phase
$\varphi_{\ell'\ell}=(\varphi_\ell'+\varphi_\ell)/2$ of the relative potential modulation
$w_{\ell'\ell}^\text{dr}(t)$.

\subsubsection{\label{sec:msl}Moving-secondary-lattice scheme}
Experimentally a site-dependent driving phase $\varphi_\ell$ has been achieved by combining two 
slightly detuned laser waves $\propto\exp(i\bk_{1,2}\cdot\br-\omega_{1,2}t)$, which create a 
shallow secondary lattice of depth $2K$ that moves with respect to the deep host
lattice.\footnote{This configuration resembles the one employed in an earlier proposal for the 
creation of artificial magnetic fields by \onlinecite{JakschZoller03}, which is based on Raman 
transitions between internal atomic states.} This moving secondary lattice causes sinusoidal 
potential modulations $w^\text{dr}_\ell(t)$ of frequency $\omega=(\omega_1-\omega_2)$ and 
spatially dependent driving phase $\varphi_{\ell}=\bq\cdot\br_\ell$ with $\bq=(\bk_1-\bk_2)$.
A configuration of particular interest is shown in Fig.~\ref{fig:flux}(b). For a square lattice 
$\bq=(q_x,q_y)$ is combined with a strong static potential gradient in one of the lattice 
directions, $\nu_\ell = x_\ell/d$ \cite{AidelsburgerEtAl13,MiyakeEtAl13}. According to
Eq.~(\ref{eq:JeffPA}), the resulting effective parameters for tunneling in $x$ and $y$ 
direction read 
\be\label{eq:Jraman}
J^\text{eff}_{x}=J_x\Jb_1\Big(\frac{K_x}{\hbar\omega}\Big)e^{i\bq\cdot(\br_{\ell'}+\br_\ell)/2},
\quad
J^\text{eff}_y=J_y\Jb_0\Big(\frac{K_y}{\hbar\omega}\Big),
\ee
where $K_i=2\sin(d_iq_i)K$, with $i=x,y$. Here $d_i$ and $J_i$ denote the lattice spacing and 
the tunneling parameter in both spatial directions, respectively.

The $y$-dependent part $q_y y$ of the Peierls phase for ``photon'' assisted tunneling in $x$ 
direction, gives rise to an effective flux of 
\be
\Phi_\text{eff}= d q_y,
\ee
piercing every lattice plaquette. The effective Hamiltonian describes particles on a square 
lattice subjected to a homogeneous magnetic field and corresponds to the paradigmatic Harper 
Hamiltonian \cite{Harper55}. It is famous for the fractal structure of its single-particle 
spectrum plotted versus $\alpha=\Phi_\text{eff}/2\pi$, the Hofstadter butterfly
\cite{Hofstadter76}. It results from the possibility that the area $d_xd_y/\alpha$ of the 
magnetic unit cell can become an incommensurate multiple of the area $d_xd_y$ of the the
square-lattice unit cell. If the potential gradient results from a Zeeman field the sign of the 
magnetic flux will depend on the spin state of the atoms \cite{AidelsburgerEtAl13,MiyakeEtAl13, 
KennedyEtAl13}. If the potential gradient is replaced by an optical superlattice, so that
$\nu_\ell=[1+(-1)^{x_\ell/d}]/2$ gives rise to staggered potential off-sets
$\nu_{\ell'\ell}\hbar\omega$, and a staggered pattern of fluxes $\pm|\Phi_\text{eff}|$
\cite{AidelsburgerEtAl11}. In order to realize the Harper Hamiltonian also for the 
superlattice configuration, the flux can be rectified by combining two moving secondary 
lattices such that driving phases $\varphi_{\ell'\ell}$ are obtained that compensate 
the staggered potential offsets \cite{AidelsburgerEtAl15}.

Experimentalists have investigated the ground state of a weakly interacting Bose gas in 
effective lattice models created using the moving-secondary-lattice scheme. Their observations 
reflect the fact that finite plaquette fluxes introduce frustration into the tunneling 
kinetics. This frustration results from the fact that the phase-winding of the wave function 
around the plaquette has to be an integer multiple of $2\pi$, while the optimal phase 
differences at the tunneling bonds, which are given by $\theta^\text{eff}_{\ell'\ell}$, sum up 
to $\Phi_\text{eff}$. The frustration becomes maximum for the maximum phase mismatch
$\Phi_\text{eff}\!\mod 2\pi=\pi$. For a fixed flux the degree of frustration can be controlled 
by the relative strength of different tunneling parameters, since it becomes energetically less 
costly to accommodate a greater share of the phase mismatch at weaker tunneling links. 
In the experiment by \onlinecite{AidelsburgerEtAl11} the bosonic ground state for a staggered 
flux configuration with $\Phi_\text{eff}=\pm\pi/2$ was explored. The unit cell contains two 
sites, giving rise to two bands. If the ratio $\gamma=|J_x^\text{eff}|/|J_y^\text{eff}|$ 
becomes larger than the critical values of $\gamma_c=\sqrt{2}$, the central minimum of the 
lowest band splits continuously into two minima that separate in $k_y$-direction. That is, 
while for $\gamma\le\gamma_c$ the wave function does not adapt its momentum to the plaquette 
flux, for larger $\gamma$ a two-fold degenerate spiral phase pattern around the lattice 
plaquettes becomes more favorable. This resembles the case of the frustrated triangular 
lattice reviewed in the previous section. However, different from the triangular lattice, a 
homogeneous density distribution, as it is favored by the repulsive interactions, is achieved 
by a coherent superposition of both ground states \cite{MoellerCooper10}, which has been 
observed in the experiment. 

Reducing the geometry to one-dimensional ladders with constant plaquette flux $\pi/2$ 
by suppressing tunneling at every other link in $x$ direction for the model of
Fig.~\ref{fig:flux}(b), the transition at $\gamma=\gamma_c$ finds an appealing interpretation 
as an analog of the Meissner effect in superconductors \cite{OrignacGiamarchi01}. For
$\gamma<\gamma_c$, corresponding to the phase of low magnetic fields, the wave function is 
stiff, so that the Peierls phases cause a circular Meissner current around the whole ladder. In 
turn, when $\gamma>\gamma_c$, the wave function adapts to the field, corresponding to the 
formation of vortices. This effect was observed by \onlinecite{AtalaEtAl14}. 

Very recently, the bosonic superfluid ground state, or more precisely a low-entropy state close 
to it, was prepared using the tilted-lattice configuration giving rise to a plaquette of $\pi$
[as depicted in Fig.~\ref{fig:flux}(b), but with $\Phi_\text{eff}=\pi$ \cite{KennedyEtAl15}]. 
Even in the presence of a rather deep lattice (of more than ten recoil energies) in the 
perpendicular $z$ direction, which leads to a significant increase of interactions and which 
can be used to reduce the dynamics to two dimensions, rather large coherence times were 
observed. This is a promising step towards the preparation of strongly correlated fractional 
quantum-Hall-type states in optical lattices with artificial magnetic fields. 

Also dynamical signatures of artificial magnetic fields have been probed experimentally. 
Conceptually maybe the most straightforward signature is the observation of the cyclotron-type 
dynamics of a single particle on an isolated plaquette of the square lattice
\cite{AidelsburgerEtAl11,AidelsburgerEtAl13}. An intriguing effect is, moreover, the 
observation of a quantized Hall velocity with \emph{thermal bosons} in a square lattice 
with homogenous flux \cite{AidelsburgerEtAl15}. For a quarter of a flux quantum per plaquette  
the elementary lattice cell is enlarged to the area $A_m=4d^2$ of four plaquettes pierced by 
one flux quantum and the Hubbard model describes four Bloch bands. The lowest band has 
favorable properties. It is rather flat, i.e.\ it is separated by a large energy gap of about 
seven times the band width, and characterized by a Chern number $C_0=1$. For the $b$th band of 
a two-dimensional lattice this topological index is defined like
\be\label{eq:chern}
C_b =\frac{1}{2\pi} \int_{\text{BZ}}\! \rd k_x\rd k_y\, \Omega_b(\bk) ,
\ee
where the integral is taken over the reduced first Brillouin zone of area $(2\pi)^2/A_m$
corresponding to the enlarged lattice cell $A_m$. Moroever, $\Omega_b(\bk)$ denotes the Berry 
curvature with respect to quasimomentum $\bk$. It is given by
$\Omega_b(\bk) = {\bm e}_z\cdot {\bf \Omega}_b(\bk)$ with
${\bm\Omega}_b(\bk)=\nabla_\bk\times{\bm A}_b(\bk)$ and Berry connection 
${\bm A}_b(\bk)=i\la u_b(\bk)| \nabla_\bk |u_b(\bk)\ra$. Here $|u_b(\bk)\ra$ is the spatially 
periodic part of the Bloch state with quasimomentum $\bk$ of band $b$. The Chern number is 
quantized and can take integer values only. This is a consequence of the fact that it 
corresponds to $(2\pi)^{-1}$ times the Berry phase associated with a closed surface in 
quasimomentum space, namely the torus given by the first-Brillouin zone. In the presence of a 
homogeneous force $\bF$ the velocity associated with a Bloch state is given by 
(see, e.g.\ \onlinecite{DiEtAl10})
\be\label{eq:vk}
{\bm v}_b(\bk) = \frac{1}{\hbar}\big[\nabla_\bk\varepsilon_b(\bk)
		-{\bm F}\times{\bm \Omega}_b(\bk)\big].
\ee
The ``anomalous'' second term describes a Hall drift \cite{KarplusLuttinger54}. For a 
homogeneously filled band, the mean velocity of the particles is given by
\be
\bar{\bm v}(\bk)=\frac{A_m}{(2\pi)^2}\int_\text{BZ} \! \rd k_x\rd k_y\,{\bm v}_b(\bk)=
-A_m\frac{C_b}{h}{\bm e}_z\times{\bm F},
\ee
where we have used Eq.~(\ref{eq:vk}) and the fact that the first term on the left-hand-side of 
Eq.~(\ref{eq:vk}) averages to zero. The velocity is proportional to the Chern number $C_b$ 
and, therfore, quantized. For a fermionic band insulator, with the $B$ lowest bands filled 
completely, this result implies a quantized Hall conductivity $\sigma_{h}=C/h$ with
$C=\sum_{b\le B}C_b$ \cite{ThoulessEtAl82}.\footnote{$\sigma_h=C e^2/h$ for charged particles 
with elementary charge $e$.}. For $C\ne0$ this is the integer quantum Hall effect and the 
system is a topological insulator called Chern insulator \cite{HasanKane10,QiZhang11}. However, 
in the experiment by \onlinecite{AidelsburgerEtAl15}, the flatness of the lowest band is 
exploited in order to create a thermal state, where weakly interacting bosons occupy (to good 
approximation) the lowest band in a uniform fashion, but no excited bands. Extracting the Hall 
displacement of the cloud in response to a force ${\bm F}$ the Chern number $C_0$ was 
measured to be $0.99(5)$ in excellent agreement with theory. Here an effective force
$F_x=\delta/d$ results from a slight detuning $\delta$ between the driving frequency and the 
potential off set between neighboring sites to be overcome by ``photon''-assisted tunneling
[Eq.~(\ref{eq:Delta})].) Also the breakdown of the Hall response was observed at a topological 
transition to a lattice structure with $C_0=0$ induced by a superlattice potential was 
observed. 

\subsubsection{\label{sec:als}Asymmetric-lattice-shaking scheme}
A different scheme for the creation of artificial magnetic fields does not require potentials 
off-sets with finite $\nu_{\ell'\ell}$, and can be realized by means of lattice shaking
\cite{StruckEtAl12,HaukeEtAl12b,StruckEtAl13}. It is based on breaking both the local 
reflection symmetry (\ref{eq:lr}) and the shift symmetry (\ref{eq:shift}) by employing
non-sinusoidal driving functions $w^\text{dr}_\ell(t)$. The fact that the optical-lattice 
physics happens at rather low energy scales, so that the driving frequencies $\omega/2\pi$ 
required for the high-frequency approximation (\ref{eq:Heff}) and (\ref{eq:Jeff_ell}) are in 
the lower kilohertz regime, allows for the implementation of practically arbitrary shaking 
functions. In a first experiment, a one-dimensional lattice has been subjected to the inertial 
force depicted in Fig.~\ref{fig:flux}(c), which led to the complex tunneling parameter 
$J_\text{eff}=|J_\text{eff}|e^{i\theta_\text{eff}}$ shown in the same figure. 
While in a one-dimensional chain no magnetic field is created by a finite Peierls phase, 
its impact, a shift of the effective dispersion relation $\varepsilon_\text{eff}(k_x)$ by
$\theta_\text{eff}/d$, can still be observed. When $\theta_\text{eff}$ (representing the
$x$ component of a homogeneous, but time-dependent vector potential) is rapidly ramped up 
this creates a significant conservative force (the shifted dispersion relation possesses a 
finite group velocity at the initial condensate momentum $k_x=0$). This initiates an 
oscillatory dynamics in the trap. When $\theta_\text{eff}$ is switched on slowly the trapped 
condensate follows the minimum of the dispersion relation. This gives rise to a peak shift by 
$k_\text{shift}=\theta_\text{eff}/d$ in the measured momentum distribution (see discussion in 
the last paragraph of section~\ref{sec:dl}), from which the Peierls phase $\theta_\text{eff}$
was inferred (data points in Fig.~\ref{fig:flux}(c)].

Asymmetric lattice shaking can be employed to realize effective magnetic fluxes through such 
lattice plaqettes that do not feature pairwise parallel edges (whose contribution to the flux 
would mutually cancel). This has been used in an experiment by \onlinecite{StruckEtAl13} to 
create a staggered flux configuration in a triangular lattice (of one-dimensional tubes), with 
fluxes
\be
\Phi^\text{eff}_\bigtriangleup=\Phi_\text{eff} \qquad\text{and}\qquad
\Phi^\text{eff}_\bigtriangledown=-\Phi_\text{eff}
\ee
for the two types of plaquette orientations. Unless $\Phi_\text{eff}=0$ or $\pi$, these fluxes 
break time-reversal symmetry in the approximate effective Hamiltonian (\ref{eq:Heff}). But they 
do not break the translational symmetry of the lattice so that the tight-binding model still 
gives rise to a single band. This is different compared to the square lattice with homogeneous 
flux [Fig.~\ref{fig:flux}(b)] realized using the moving-secondary-lattice scheme, where an 
enlarged magnetic unit cell leads to the formation of several bands. Another difference 
between both schemes concerns the limit of small driving amplitudes $K/\hbar\omega$: for the 
moving-secondary-lattice-assisted tunneling against a finite off-set
$\nu_{\ell'\ell}\hbar\omega$ the amplitude of the tunneling matrix element vanishes and 
the Peierls phase remains constant [like for $J_x^\text{eff}$ in Eq.~(\ref{eq:Jraman})], 
whereas for the asymmetric driving scheme with $\nu_{\ell'\ell}=0$ the amplitude remains 
finite and the Peierls phase continuously approaches zero [as shown in Fig.~\ref{fig:flux}(c)]. 

The possibility of the asymmetric driving scheme to continuously tune Peierls phases and 
plaquette fluxes \emph{in situ} has been employed in the triangular lattice experiment in order 
to tune the system away from the $\pi$-flux configuration resulting from the sign-change of the 
tunneling parameter discussed in the previous section 
\ref{sec:frustration}. Realizing 
\be
\Phi_\text{eff}=\pi+\delta,
\ee
a small $\delta$ favors one of the two symmetry broken ground states of the weakly interacting 
Bose gas [Fig.~\ref{fig:tri}(c)], so that $\delta$ controls a first-order phase transition at
$\delta_c=0$. As a signature of the discontinuous nature of the transition the disfavored state 
remains metastable in the vicinity of the transition (potentially causing hysteresis). This was 
inferred from the observation that the distribution shown in Fig.~\ref{fig:tri}(c) becomes 
asymmetric for finite $\delta$, but remains bimodal up to $|\delta|\approx0.1\pi$ 
\cite{StruckEtAl13}. It has, moreover, been observed that the time-reversal symmetry breaking 
and the metastable state vanish for large temperatures. The interesting question whether
time-reversal symmetry breaking disappears together with Bose condensation or in a separate 
transition at a higher temperature could not be resolved.

\subsubsection{Further possibilities}

Apart from the moving-secondary-lattice scheme and the asymmetric-lattice-shaking scheme, there 
are also proposals for the dynamic creation of artificial magnetic fluxes that are based on a 
joint modulation of both on-site energies and tunneling parameters \cite{SoerensenEtAl05,
LimEtAl08}. Moreover, extensions of the moving-secondary-lattice and the asymmetric shaking 
schemes have been proposed for the creation of Haldane-type hexagonal-lattice Chern 
insulators, non-abelian gauge fields (spin-orbit coupling) and topological (spin-Hall) 
insulators, as well as Weyl semi-metals \cite{BermudezEtAl12, StruckEtAl12, KennedyEtAl13, 
BaurEtAl14, DubcekEtAl14}. Another approach for Floquet engineering of artificial 
magnetic fields or spin-orbit coupling is to construct driving protocols given by a sequence 
of pulses during which different external fields are present \cite{GoldmanDalibard14}. This 
includes proposals for the creation of artificial magnetic fields \cite{SoerensenEtAl05, 
CreffieldSols14, CreffieldEtAl16} and of spin-orbit coupling  in continuous (non-lattice) 
systems of ultracold atoms using a sequence of magnetic field pulses \cite{XuEtAl13, 
AndersonEtAl13} as well as for the realization of Floquet topological Chern insulators using 
pulsed tunneling matrix elements in lattice systems \cite{KitagawaEtAl11, RudnerEtAl13}. The 
latter two references can be viewed as variants of schemes known under the label \emph{``
Floquet-topological insulators''}. These are based on intermediate driving frequencies and 
will be discussed in section \ref{sec:fti} as well as in section \ref{sec:EdgeStates}, which 
is devoted to anomalous topological edge states in periodically driven lattice systems. 
Finally, we would like to mention two proposals for the dynamic creation of artificial gauge 
fields not relying on standard optical lattices. The first one suggests to use periodically 
modulated spin-dependent optical potentials in order to effectively engineer lattices with sub-
wave-length spacing featuring bands with non-zero Chern numbers \cite{NascimbeneEtAl15}. The 
second one is based on the coherent resonant coupling of the eigenstates of a strong harmonic 
confinement, playing the role of lattice sites in a ``synthetic dimension'' \cite{PriceEtAl16},
in analogy to earlier work where internal atomic states where used for this purpose 
\cite{BoadaEtAl12,CeliEtAl14,ManciniEtAl15,StuhlEtAl15, PriceEtAl16}.

It is an interesting perspective to combine the two schemes discussed in this section with 
strong interactions. The flat topological band of the $\pi/2$ Harper model realized by
\onlinecite{AidelsburgerEtAl15} together with the promising creation of low-entropy states in 
such a system \cite{KennedyEtAl15} make this system a candidate for the stabilization of 
topologically ordered fractional-quantum-Hall-type states (fractional Chern insulators, 
\onlinecite{BergholtzZhao13, ParameswaranEtAl13}). Moreover, in the hard-core boson limit, the 
time-reversal symmetry breaking, as it can be induced by asymmetrically shaking the triangular 
lattice, introduces Dzyaloshinskii-Moriya interactions
${\bm D}\cdot(\hat{\bm S}_{\ell'}\times \hat{\bm S}_{\ell})$ to the effective spin model
(\ref{eq:Hspin}), extending the tool box for quantum engineering of spin Hamiltonians. Namely, 
for hard-core bosons the tunneling term $-(J_\text{eff}\,\aa_{\ell'}\ao_\ell + \text{h.c.})$ 
with complex $-J_\text{eff}=R+iI$ corresponds to 
\be
2R \Big(\hat{S}_{\ell'}^{x}\hat{S}_{\ell}^{x}+\hat{S}_{\ell'}^{y}\hat{S}_{\ell}^{y}\Big)
+2I {\bm e}_z\cdot\Big(\hat{\bm S}_{\ell'}\times \hat{\bm S}_{\ell}\Big)
\ee
in the language of spin-1/2 operators.

\subsection{Coherent resonant band coupling}

\begin{figure}[t]
\includegraphics[width=1\linewidth]{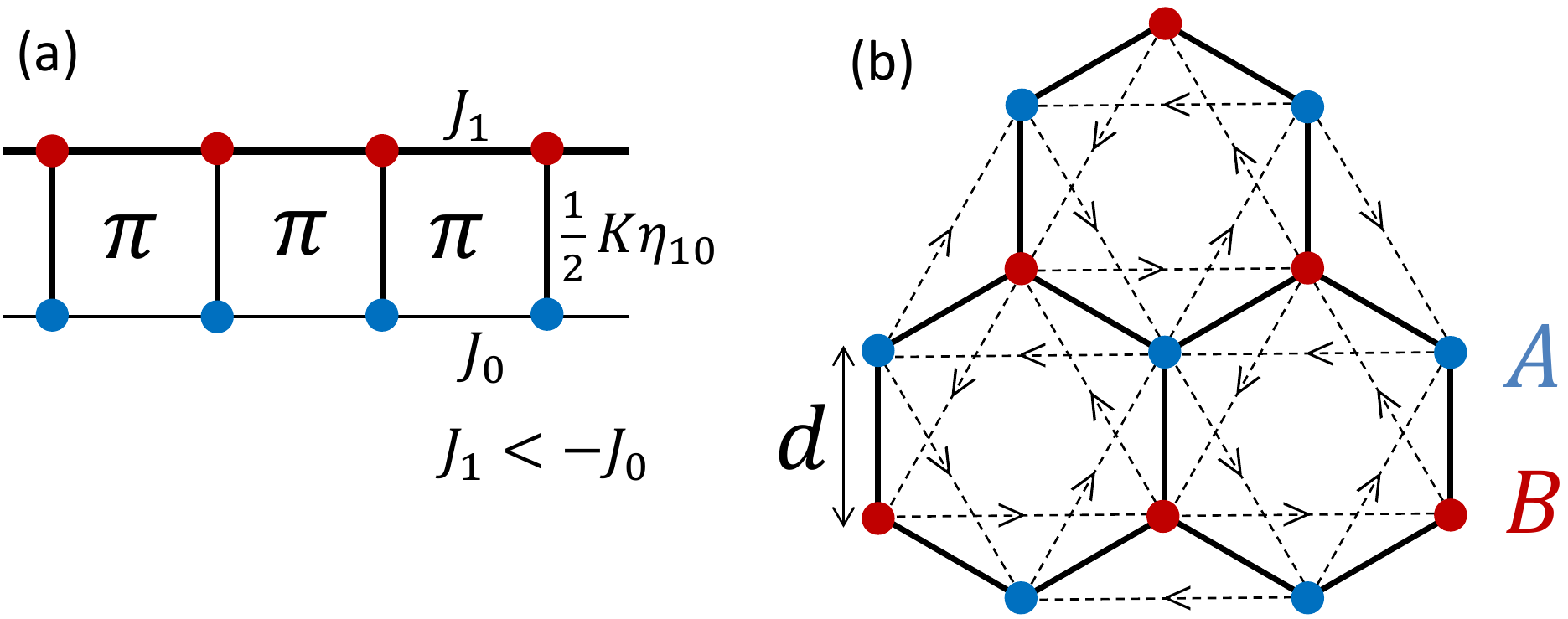}
\centering
\caption{\label{fig:BandHex} (a) By resonantly coupling the two lowest bands of a cosine 
lattice a frustrated ladder is created with plaquette fluxes of $\pi$. (b) Hexagonal lattice. 
The effective Hamiltonian of the driven system including next-nearest-neighbor tunneling with 
Peierls phase $\theta$ (along the dashed lines).} 
\end{figure}

Periodic driving can not only bridge large energy off-sets between neighboring sites, but it 
can also induce coherent resonant coupling to excited Bloch bands. This possibility has been 
explored experimentally in Chu's group in Stanford \cite{GemelkeEtAl05}, in Greiner's group at 
Harvard \cite{BakrEtAl11}, and in Chin's group in Chicago \cite{ParkerEtAl13, HaEtAl15}. 

In the Stanford and Chicago experiments, conducted with weakly interacting bosons, the lowest 
two bands of a cosine lattice have been coupled by means of lattice shaking with the driving 
frequency $\hbar\omega$ bridging the gap between both bands. Assuming that the coupling to even
higher-lying bands is off-resonant and negligible, in this way an effective hybridized band 
structure is created. The coupling between both bands happens predominantly on-site. 
Introducing the label $\alpha=0,1,\ldots$ for the Bloch bands and their Wannier orbitals with 
respect to one lattice direction, $x$, an inertial force (\ref{eq:Fin}) oriented in this 
direction gives rise to oscillating coupling matrix elements
$K\eta_{\alpha'\alpha}\cos(\omega t)\aa_{\alpha'\ell}\ao_{\alpha\ell}$. Here the dimensionless 
dipole matrix element $\eta_{\alpha'\alpha}$ vanishes for Wannier states of the same parity, 
i.e.\ when $(\alpha'-\alpha)$ is even. Analogously to the case of photon-assisted tunneling 
(\ref{eq:Jeffamp}) via a modulation of the tunneling matrix element (\ref{eq:Jamp}), one finds 
an effective description where the $\alpha=1$ Wannier states have the shifted energy
$\epsilon_1-\hbar\omega$ and couple to the $\alpha=0$ states via effective matrix element
$K\eta_{10}/2$. The band-coupled system can be viewed as a ladder, with the Wannier states of 
each band forming one leg [Fig.~\ref{fig:BandHex}(a)]. The tunneling parameters $J_\alpha$ of 
both bands differ in sign and magnitude, $J_1<-J_0$, so that the ladder is frustrated by 
plaquette fluxes of $\pi$ \cite{StraeterEckardt15}. The effective band structure results from 
hybridization of the effective dispersion relations of both bands, 
$\varepsilon_0^\text{eff}(k_x)=-J_0\Jb_0(K/\hbar\omega)\cos(dk_x)$ and
$\varepsilon_1^\text{eff}(k_x)=\epsilon_1-\hbar\omega + J_1\Jb_0(K/\hbar\omega)\cos(dk_x)$, 
where $\epsilon_1$ is the band-center energy of the first excited band \cite{DreseHolthaus96}. 

In the experiment by \onlinecite{GemelkeEtAl05}, $\varepsilon_0^\text{eff}(0)$ was tuned to 
resonance with $\varepsilon_1^\text{eff}(\pi/d)$ and coherent oscillations between both states, 
attributed to scattering, were observed after the forcing was switched on suddenly. In the 
Chicago experiments, tuning $\epsilon_1-\hbar\omega<0$ a hybridized 
band with two inequivalent minima $k_x=\pm q$ was created. Similar like in the case of the 
kinetically frustrated triangular lattice, repulsive interactions favor Bose condensation in 
one of the two minima, but not in both. Like in a ferromagnet, the experimentalists observed 
the formation of spatial domains with $+q$ or $-q$ correlations \cite{ParkerEtAl13}. The domain 
size was controlled by how fast the driving was switched on, with large domains obtained 
for slow ramps. In a subsequent experiment Bragg spectroscopy was used to measure the 
dispersion relation of the elementary Bogoliubov excitations of the system condensed into one 
of the minima \cite{HaEtAl15}. It is phonon-like near the condensate momentum and can feature a 
local minimum at the second minimum of the effective dispersion relation of the non-interacting 
gas (see also \onlinecite{StruckEtAl13}). This structure reminds of a roton minimum resulting 
from long-ranged interactions.

The Harvard experiment \cite{BakrEtAl11} was performed in a rather deep lattice, where 
interactions are strong and band coupling can be understood on the level of a single site. By 
employing a modulation of the lattice depth in one direction, the lowest-band Wannier orbital 
was coupled to states of the same parity in this direction (see, e.g.\
\onlinecite{LackiZakrezewski13}).\footnote{Resonant transitions into orbital states of 
opposite parity can in principle also occur in the presence of interactions, namely when two 
particles jointly scatter into the excited state \cite{Sowinski12} or in the form of
density-induced orbital-changing tunneling processes, as they have recently been shown to give 
rise to exotic model systems \cite{BiedronEtAl16, DuttaEtAl15b, PrzysieznaEtAl15}.} The 
resonance frequency for coupling to the second excited state ($\alpha=2$) was found to depend 
crucially on the on-site occupation of both states as a consequence of strong orbital-
dependent interactions. This allows to engineer number-selective adiabatic passages, where a 
single particle is transferred to the excited band. In particular, by slowly ramping down the 
driving frequency, a sequence of such processes subsequently transfers all, but a single 
particle to the excited band, irrespective of the initial occupation. In this way an 
algorithmic cooling procedure was implemented: A state with an arbitrary number of atoms $\ge1$
in the lowest Wannier orbital on every site is eventually transformed into a state with one 
atom in the lowest Wannier orbital per site. Entropy has been transferred to an excited band, 
from where it can be removed by selectively taking away the excited atoms.

\subsection{\label{sec:fti}Floquet topological insulators}

All the experiments discussed here so far, rely on high-frequency forcing, where $\hbar\omega$ 
is large compared to the tunneling matrix elements. This is different for a class of recent 
proposals for the Floquet engineering of lattice systems with topologically non-trivial band 
structures \cite{OkaAoki09,KitagawaEtAl10,LindnerEtAl11,CayssolEtAl13}. These schemes, known as 
\emph{Floquet topological insulators}, rely on driving frequencies that are only moderately 
larger than the tunneling matrix elements. The prototype of a Floquet topological insulator was 
originally proposed for graphene irradiated with circularly polarized light \cite{OkaAoki09}. 
It is based on the observation that a hexagonal tight-binding lattice subjected to a circular 
force (\ref{eq:Fcirc}) possesses an effective band structure with a gap separating two bands 
with opposite Chern numbers $\pm1$. Recently, this Floquet topological band structure has 
been realized and probed in two different experimental platforms. In a first experiment with 
photons in a hexagonal array of optical wave guides, the chiral transport of localized 
particles at the boundary of the system was observed \emph{in situ} \cite{RechtsmanEtAl13}. 
This is a signature of the chiral edge states related to Chern bands via the bulk-boundary 
correspondence \cite{HasanKane10,QiZhang11}. The second experiment was conducted by
\onlinecite{JotzuEtAl14} in Esslinger's group in Zurich with ultracold fermionic atoms in a 
shaken hexagonal-like brick-wall lattice. Here a finite Hall conductivity of the bulk system 
was measured. 

A hexagonal lattice with isotropic nearest-neighbor tunneling [Fig.~\ref{fig:BandHex}(b), solid 
lines] subjected to circular forcing~(\ref{eq:Fcirc}) is described by the Hamiltonian
(\ref{eq:Hprime}) with time-dependent Peierls phases
$\theta_{\ell'\ell}(t)=K\sin(\omega t-\varphi_{\ell'\ell})$. Here $K=Fd$ and the driving phase
$\varphi_{\ell'\ell}$ is directly determined by the spatial direction of tunneling. In the
high-frequency limit $\hbar\omega\gg J$, $\Ho'(t)$ can be approximated by its time average 
giving rise to effective tunneling matrix elements (\ref{eq:Jeff}) between nearest neighbors. 
However, if the frequency is lowered, also the second-order term $\Ho_F^{(2)}$ in the
high-frequency expansion (\ref{eq:hf}) becomes relevant. Its contribution to the effective 
Hamiltonian $\Ho_F$ results from processes where a particle tunnels twice during one driving 
period. Using Eq.~(\ref{eq:Hhf}) with
$\Ho_m=-\sum_{\la\ell'\ell\ra}J\Jb_m(K/\hbar\omega)\aa_{\ell'}\ao_\ell$, in second order one 
finds the kinetics to be described by the approximate effective Hamiltonian
\cite{KitagawaEtAl11} 
\bes\label{eq:Hfti}
\Ho_F&\approx&\Ho^{\prime(1)}_F+\Ho^{\prime(2)}_F
\nonumber\\
&=&-J^{(1)}_\text{eff}\sum_{\la\ell'\ell\ra}\aa_{\ell'}\ao_\ell
	-J_\text{eff}^{(2)} \sum_{\lla\ell'\ell\rra} 
		e^{-i\sigma_{\ell'\ell}\theta} \aa_{\ell'}\ao_\ell .
\ees
Here $J^{(1)}_\text{eff}=J\Jb_0(K/\hbar\omega)$, 
$J_\text{eff}^{(2)}\simeq\sqrt{3}[J\Jb_1(K/\hbar\omega)]^2/\hbar\omega$ (neglecting terms with
$m\ge2$), $\theta=\pi/2$ and next-nearest-neighbor pairs
$\lla\ell'\ell\rra$ with $\sigma_{\ell'\ell}=1$ (-1) for tunneling in anticlockwise (clockwise) 
direction around a hexagonal plaquette.\footnote{Within the second-order Floquet-Magnus 
expansion of the Floquet Hamiltonian (\ref{eq:FM}) the amplitude of the next-nearest-neighbor 
tunneling matrix elements depends on the direction of tunneling, so that the single-particle
band-structure breaks the discrete rotational symmetry of the hexagonal lattice. This is an 
artifact of the approximation related to the fact that the driving phase in $x$ and $y$ 
direction differs by $\pi/2$ for circular forcing \cite{EckardtAnisimovas15}. It illustrates 
the discussion following Eq.~(\ref{eq:H_FM}).} This model is the paradigmatic Haldane model, 
the prototype of a topological Chern insulator \cite{Haldane88}. For finite next-nearest 
neighbor tunneling matrix elements the band structure acquires a gap separating two bands of 
opposite Chern number $\pm1$. By introducing an energy difference $\Delta_{AB}$ between both 
sublattices $A$ and $B$ [Fig.~\ref{fig:BandHex}(b)], the Chern numbers vanish when at
$|\Delta_{AB}|=\Delta_{AB}^{(c)}$ the band gap closes. $|\Delta_{AB}^{(c)}|$ is maximum,
$\approx2.6|J_\text{eff}^{(2)}|$, for $\theta=\pm\pi/2$ and vanishes for $\theta =0,\pi$. 

In the Zurich experiment a distorted optical hexagonal lattice (a brick-wall lattice) with a 
tunable off-set $\Delta_{AB}$ is filled with spin-polarized (i.e.\ non-interacting) fermions 
and an elliptical force $\bF(t)=F[\cos(\omega t){\bm e}_x+\cos(\omega t-\varphi){\bm e}_y]$ is 
applied via lattice shaking (\onlinecite{JotzuEtAl14}, see also \onlinecite{ZhengZhai14}). The 
system is described by an anisotropic effective model, whose phase diagram resembles that of 
the Haldane model with $\varphi$ playing a role similar to $\theta$. This phase diagram is 
mapped out by measuring the Hall response of the system.

\subsection{Floquet engineering of interactions}
In most of the experiments described above, periodic forcing was employed to effectively modify 
the single-particle Hamiltonian of the system, describing tunneling between neighboring lattice 
sites or the coupling between different Bloch bands. In contrast, the on-site interactions 
among the particles were not altered significantly by the driving. Exceptions are given by the 
experiments of Greiner's group discussed above. Here interactions were strong enough to shift 
the resonance condition for ``photon''-assisted processes, tunneling \cite{MaEtAl11, ChenEtAl11}
or band coupling \cite{BakrEtAl11}, so that they became occupation-number selective 
(effectively realizing an extended Hubbard model, \onlinecite{DuttaEtAl15}). In this way 
tunneling or band coupling are not described by single-particle terms in the effective 
Hamiltonian anymore, which are quadratic in the annihilation and creation operators, and must 
be viewed as a form of interactions. 

It is an interesting prospect to combine such a technique with driving schemes for the creation 
of artificial gauge fields as we discussed them in Sec.~\ref{sec:HFgauge}
\cite{BermudezPorras15,StraeterEtAl16,CardarelliEtAl16}. For that purpose one has to consider a 
lattice system, where the strong energy off-sets that have to be overcome by resonant
``photon''-assisted tunneling are determined not only by static external potentials, but also 
by the strong interactions among the particles. 
Choosing the resonance condition $U=\nu\hbar\omega+\delta U$, with $|\delta U|\ll\hbar\omega$ 
between the Hubbard parameter for on-site interactions and the driving frequency, tunneling 
from site $\ell$ to site $\ell'$ corresponds to a potential energy change of an integer number
$\nu_{\ell'\ell}$ of quanta $\hbar\omega$. Considering, e.g., spinless bosons, from
Eq.~(\ref{eq:Hint}) one obtains $\nu_{\ell'\ell} 
= \nu_\ell'-\nu_\ell +\nu[n_{\ell'}-(\nu_\ell-1)]$, with site occupations $n_\ell$ and
$n_{\ell'}$ before the tunneling event. For $\hbar\omega\gg J$, tunneling is suppressed for
$\nu_{\ell'\ell}\ne0$, unless it is reestablished via ``photon''-assisted tunneling, giving 
rise to number-dependent effective tunneling parameters 
$J^\text{eff}_{\ell'\ell}(n_{\ell'},n_\ell)$. The system is then described by an approximate 
effective Hamiltonian that, in rotating wave approximation, takes the form
\be
\Ho_\text{eff}
=-\sum_{\la\ell'\ell\ra} \aa_{\ell'}\ao_\ell J^\text{eff}_{\ell'\ell}(\no_\ell',\no_\ell)
+ \frac{U_\text{eff}}{2}\sum_\ell \no_\ell(\no_\ell-1).
\ee
The effective Hubbard parameter is given by the non-resonant part of the interactions,
$U_\text{eff}=\delta U=U-\nu\hbar\omega$, which is not integrated out when transforming to the 
rotating frame and whose magnitude and sign can be controlled by the driving frequency (this is 
true also for fermionic systems). 

\onlinecite{BermudezPorras15}, proposed such schemes, where the ``photon''-assisted tunneling 
is induced by a moving-secondary-lattice as described in Sec.~\ref{sec:msl}. They lead to 
effective tunneling matrix elements described by Eq.~(\ref{eq:JeffPA}), with $\nu_{\ell'\ell}$ 
replaced by an operator involving the occupation numbers of the particles. In this way they 
show, among others, how to engineer models where the magnetic field felt by one atomic species 
depends dynamically on the state of another species. Moreover, \onlinecite{StraeterEtAl16} 
describe how to realize the physics of one-dimensional lattice anyons by inducing
photon-assisted tunneling via asymmetric lattice shaking (see Sec.~\ref{sec:als}).
They use a mapping of the anyons to bosons with number-dependent tunneling parameters
$J_{\ell+1,\ell}= |J|e^{i\theta\no_{\ell+1}}$ and $J_{\ell-1,\ell}= |J|e^{-i\theta\no_{\ell}}$, 
which had been exploited already in a previous proposal based on Raman-assisted tunneling
\cite{KeilmannEtAl10}. Finally, \onlinecite{CardarelliEtAl16}, propose a scheme based on the 
fact that ``photon''-assisted tunneling induced by a sinusoidal modulation of the tunneling 
matrix element (as they can be induced by a modulation of the lattice depth) gives rise to 
single-``photon'' transitions only. This can be seen from that fact that Eq.~(\ref{eq:Jeffamp}) 
describes non-zero effective tunneling matrix elements for $|\nu|\le1$ only. Thus, 
superimposing sinusoidal modulations at different frequencies, one can individually address 
tunneling processes corresponding to different number-dependent energy off-sets, with the 
amplitude and the phase of the effective tunneling matrix elements directly corresponding to 
the amplitude and the phase of the modulation. 

An alternative approach for achieving number-dependent tunneling matrix elements consists in a 
modulation of the interaction strength \cite{GongEtAl09}, as it can be achieved in a 
system of ultracold atoms by employing a magnetic Feshbach resonance, giving rise to a
time-dependent Hubbard parameter $U(t)=U_0+U_\text{dr}(t)$ with
$U_\text{dr}(t+T)=U_\text{dr}(t)$ and $\int_0^T\!\rd t\, U_\text{dr}(t)=0$. In this way the 
energy of multiply occupied lattice sites is modulated in time, so that for spinless bosons 
tunneling from $\ell$ to $\ell'$ is connected to an energy change of
$U(t)[n_\ell'-(n_\ell-1)]$. Thus, for $U_0\ll J$ and sinusoidal forcing
$U_\text{dr}(t)=U_1\cos(\omega t)$, the effective tunneling matrix elements
$J^\text{eff}_{\ell'\ell}(n_{\ell'},n_\ell)$ are number dependent and given by
Eq.~(\ref{eq:Jeff}) with $K$ replaced by $K_{\ell'\ell}=U_1[n_\ell'-(n_\ell-1)]$. Signatures 
of this effect, which was first described theoretically by \onlinecite{GongEtAl09}, have 
recently been observed experimentally in N\"agerl's group \cite{MeinertEtAl16} by measuring 
the number of multiply occupied sites after a quench. It is moreover, proposed to use this 
principle to engineer exotic many-body states of matter \cite{RappEtAl12, GreschnerEtAl14, 
DiLibertoEtAl14, DuttaEtAl16} and to realize disordered tunneling matrix elements by modulating the 
interactions with randomly distributed localized atoms of a second species \cite{KosiorEtAl15}. 

Another driving-induced modification of the interactions is described by higher-order 
corrections of the effective Hamiltonian in high-frequency approximation (\ref{eq:Hhf}). If the 
Hubbard interactions are time-independent, so that they contribute to the Fourier component
$\Ho_0$ of the Hamiltonian only, the leading correction involving the interactions appears in 
the third-order term $\Ho_F^{(3)}$.\footnote{Within the Floquet-Magnus expansion of the Floquet 
Hamiltonian (\ref{eq:FM}) an interaction correction $\propto JU/(\hbar\omega)$ appears already 
in the second-order term \cite{VerdenyEtAl13,BukovEtAl15} [$\Ho_{t_0}^{F(2)}$ in
Eq.~(\ref{eq:H_FM}) contains $\Ho_\text{int}$ through $\Ho_0$]. However, this correction 
results from the expansion of the unitary micromotion operator and thus does not alter the 
spectrum on the order of $JU/(\hbar\omega)$ (see discussion below Eq.~(\ref{eq:H_FM})).} It 
reads \cite{EckardtAnisimovas15}
\be
\Ho_F^{(\text{int},3)} 
	= \sum_{m\ne0}\frac{\big[\Ho_{-m},\big[\Ho_\text{int},\Ho_m\big]\big]}{2(m\hbar\omega)^2},
\ee
where $\Ho_m$ denote the Fourier components of the single-particle Hamiltonian, so that 
$\Ho_F^{(\text{int},3)}\propto UJ^2/(\hbar\omega)^2$. When $U\gg J$, this term can matter in 
Floquet topological systems (Sec.~\ref{sec:fti}), where the frequency is moderately large 
only, such that effective next-nearest-neighbor tunneling matrix elements
$J_\text{eff}^{(2)}\propto J^2/(\hbar\omega)$ [Eq.~\ref{eq:Hfti}] play a crucial role. 
For the model of Sec.~\ref{sec:fti}, the $\Ho_m$ are specified above Eq.~(\ref{eq:Hfti}).
The effect of $\Ho_F^{(\text{int},3)}$ is to ``smear-out'' the interactions. It creates 
effective nearest-neighbor interactions
$\frac{V_\text{eff}}{2}\sum_{\la\ell'\ell\ra}\no_{\ell'}\no_\ell$ (at the cost of lowering the 
on-site interactions to $U-zV_\text{eff}/2$ with coordination number $z$) and also gives rise 
to density-assisted and two-particle tunneling \cite{EckardtAnisimovas15}. Numerical studies 
based on the exact diagonalization \cite{AnisimovasEtAl15,RaciunasEtAl16} suggest that this 
smearing out tends to have a negative impact on the possible stabilization of fractional-Chern-
insulator states in Floquet topological band structures recently proposed by
\onlinecite{GrushinEtAl14}. 


\section{\label{sec:FloquetPicture}The Floquet picture}
So far, we have argued that a simple high-frequency approximation provides a suitable 
description of a variety of recent experiments, where ultracold atoms in optical lattices were 
controlled by means of periodic driving. In this section we will discuss the limitations of 
this approximation and effects beyond it. This requires a treatment in terms of the extended 
Floquet Hilbert space.

\subsection{Extended Floquet Hilbert space}
By plugging the Floquet state given by Eq.~(\ref{eq:psi_nm}) into the Schr\"odinger equation
(\ref{eq:schroedinger}), one obtains 
$[\Ho(t)-i\hbar\rd_t]|u_{nm}(t)\ra=\varepsilon_{nm}|u_{nm}(t)\ra$. As was pointed out by 
\onlinecite{Sambe73}, this equation can be interpreted as the eigenvalue problem of the hermitian \emph{quasienergy operator} 
\be\label{eq:Q}
\Qo(t)=\Ho(t)-i\hbar\rd_t,
\ee 
acting in an extended Hilbert space $\mathcal{F}=\mathcal{H}\otimes\mathcal{T}$. This Floquet 
space is the product space of the physical state space $\mathcal{H}$ and the space of
time-periodic functions (with period $T$) $\mathcal{T}$. In $\mathcal{F}$ the scalar product 
combines the scalar product of $\mathcal{H}$ with time averaging and reads
\be
\lla u|v\rra =\frac{1}{T}\int_0^T\!\rd t\, \la u(t)|v(t)\ra .
\ee 
We will use a double ket $|u\rra$ for elements of $\mathcal{F}$; the corresponding state at 
time $t$ in $\mathcal{H}$ is denoted $|u(t)\ra$. Vice versa, a state $|v(t)\ra=|v(t+T)\ra$, 
including its full periodic time dependence, is written as $|v\rra$ when considered as element 
of $\mathcal{F}$. 
Likewise, an operator acting in $\mathcal{F}$ will be indicated by an overbar to distinguish it 
from operators acting in $\mathcal{H}$, which are marked by a caret. For example, $\bar{Q}$ 
denotes the quasienergy operator. Its eigenvalue problem is now written like
\be\label{eq:Qeig}
\bar{Q}|u_{nm}\rra=\varepsilon_{nm}|u_{nm}\rra.
\ee
In Floquet space $|u_{nm}\rra$ and $|u_{nm'}\rra$, defined by Eq.~(\ref{eq:eps_nm}), constitute 
independent orthogonal solutions if $m'\ne m$, so that the quasienergy spectrum is periodic 
with period $\hbar\omega$. Despite this redundancy, working in $\mathcal{F}$ has the advantage 
that one can use both methods and intuition developed for autonomous systems. 

From a complete basis of orthogonal states $|\alpha\ra$ of $\mathcal{H}$, we can construct a 
complete basis of orthogonal states $|\alpha m\rra$ of $\mathcal{F}$, given by 
$|\alpha m(t)\ra = |\alpha\ra e^{im\omega t}$ with integer $m$. In terms of these basis states 
the quasienergy operator possesses matrix elements 
\be\label{eq:Qmatrix}
\lla\alpha' m'|\bar{Q}|\alpha m\rra
	= \la\alpha'|\Ho_{m'-m}|\alpha\ra+\delta_{m'm}\delta_{\alpha'\alpha}m\hbar\omega.
\ee
The matrix possesses a transparent block structure with respect to $m$. The diagonal blocks are 
determined by the time-averaged Hamiltonian $\Ho_0$ and shifted with respect to each other by 
integer multiples of $\hbar\omega$ in quasienergy. This structure resembles that of a quantum 
system coupled to a photon-like mode in the classical limit of large photon numbers. In this 
picture $m$ plays the role of a relative photon number. The quasienergy eigenvalue problem
(\ref{eq:Qeig}) is, thus, closely related to the dressed-atom picture
\cite{AtomPhotonInteractions} for a quantum system driven by coherent radiation
\cite{EckardtHolthaus08a}. Therefore, $m$ is often called the ``photon'' number and the matrix 
elements of $\Ho_{m}$ are said to describe $m$-``photon'' processes. A unitary operator
$\bar{U}_F$ that block diagonalizes $\bar{Q}$ with respect to the ``photon'' index $m$ in
$\mathcal{F}$ corresponds directly to a time-periodic unitary micromotion operator $\Uo_F(t)$ 
in $\mathcal{H}$, as it appears in Eqs.~(\ref{eq:HFgauge}) and (\ref{eq:tri}). It produces 
diagonal blocks given by the effective Hamiltonian (\ref{eq:HFgauge}), 
\be
\lla \alpha'm'|\bar{U}_F^\dag\bar{Q}\bar{U}_F|\alpha m\rra 
= \delta_{m'm}\Big( \la\alpha'|\Ho_F|\alpha\ra +\delta_{\alpha'\alpha}m\hbar\omega\Big).
\ee

\subsection{\label{sec:highfreq}High-frequency approximation}

If $\hbar\omega$ is large compared both to the spectral width $\Ho_0$ and the matrix elements 
of the $\Ho_{m\ne0}$, states of different subspaces $m$ are energetically well separated and 
coupled only weakly to each other. It is, therefore, a reasonable approximation to neglect the 
off-diagonal blocks of the quasienergy operator (\ref{eq:Qmatrix}) and to approximate the 
effective Hamiltonian $\Ho_F$ by the time-averaged Hamiltonian $\Ho_0$. 
Corrections, resulting from the perturbative admixture of states with $m'\ne m$ to those of the 
subspace $m$, can be obtained systematically by means of degenerate perturbation theory. The 
high-frequency expansion (\ref{eq:hf}) is equivalent to such a perturbative approach, where the 
``photonic'' part $\delta_{\alpha'\alpha}\delta_{m'm}m\hbar\omega$ of the matrix
(\ref{eq:Qmatrix}) constitutes the unperturbed problem and the Hamiltonian (its Fourier 
components $\Ho_m$) the perturbation \cite{EckardtAnisimovas15}. 

The perturbative approach underlying the high-frequency expansion (\ref{eq:hf}) can be 
expected to converge only as long as the the quasienergy levels originating from different 
unperturbed subspaces $m$ remain energetically well separated. This rough statement is similar 
to the rigorous convergence criterion for the Floquet-Magnus expansion given by 
\onlinecite{CasasEtAl00} (see discussion at the end of section \ref{sec:GenProp}). In a large 
system of many particles reasonable driving frequencies $\hbar\omega$ will always be smaller 
than the spectral width of the full time-averaged Hamiltonian $\Ho_0$, so that the
high-frequency expansion cannot be expected to converge (unless the state space of the system 
is effectively divided into small subspaces due to symmetry or localization). 
Nevertheless, even in this case the high-frequency approximation (\ref{eq:Heff}) can still 
provide a suitable description of a driven many-body system on a {finite time scale}, 
provided $\hbar\omega$ is large compared to typical intensive energy scales, such as the 
tunneling parameter $J$ or the Hubbard interaction $U$ for a driven Hubbard model 
\cite{EckardtEtAl05b}. This statement is not surprising given the fact that in the previous 
section \ref{sec:experiments} we were able to explain a variety of experimental observations in 
terms of the high-frequency approximation. On longer times, deviations from the high-frequency 
approximation will, however, eventually make themselves felt as heating.

\subsection{\label{sec:heating}Heating and long-time limit}

\begin{figure}[t]
\centering
\includegraphics[width=0.9\linewidth]{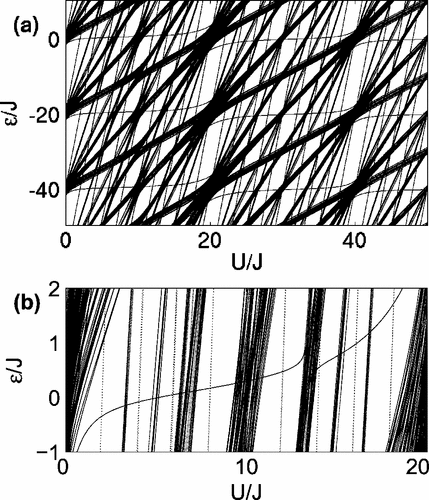}
\caption{\label{fig:spec} Quasienergy spectrum of small Bose Hubbard chain (five particles on 
five sites), subjected to a sinusoidal force of frequency $\hbar\omega/J=20$ and amplitude
$K/\hbar\omega=2$. (b) Zoom into (a). (taken from \onlinecite{EckardtHolthaus08b})} 
\end{figure}

In order to illustrate the break down of the high-frequency approximation, let us discuss a 
specific example \cite{EckardtHolthaus08b}. Fig.~\ref{fig:spec} shows the exact quasienergy 
spectrum of a small Bose-Hubbard chain subjected to a sinusoidal force of frequency
$\hbar\omega/J=20$, plotted versus the interaction strength $U/J$. For $U/J\gg1$, the spectrum 
of $\Ho_0$ consits of bands whose energies increase linearly with $U$. These bands lie above 
the horizontal ground-state level, corresponding to a Mott-insulator-like state with one 
particle localized at every lattice site. They contain states characterized by delocalized 
particle-hole excitations. The spectrum of $\Ho_0$ can clearly be identified in
Fig.~\ref{fig:spec}, as well as copies of it, shifted by $-\hbar\omega$, $-2\hbar\omega$,
\ldots. When states belonging to different copies (``photon'' numbers $m$) become degenerate, 
this leads to the formation of avoided level crossings, the size of which reflects the coupling 
strength. Near $U=\hbar\omega=20J$ and $U=2\hbar\omega=40J$, the ground state participates in a 
large avoided-crossing-like feature (involving many bands), associated with the resonant 
creation of a particle-hole pair of energy $U$. The size of this feature is determined by the 
coupling matrix element $J$ associated with the creation of a particle-hole pair. For 
$U<\hbar\omega$, a smaller avoided crossing is visible in subfigure (b) near 
$U=2\hbar\omega/3\approx 13 J$. It can be attributed to the creation of two coupled
particle-hole excitations of energy $3U$ in a two-``photon'' process. The size of the avoided 
crossing, which reflects the coupling matrix element, is of the order of
$\sim J^2/(\hbar\omega)$. Whereas the numerator of this factor results from the fact that two 
tunneling processes are required to create a two particle-hole pairs, the denominator indicates 
that the transition occurs via intermediate states (having a single-particle-hole pair) that 
are separated by a large energy $\sim\hbar\omega$ \cite{EckardtAnisimovas15}. For even smaller 
values of $U$, the $m=0$ ground state will cross even higher lying bands of the $m<0$ copies, 
which contain states with three and more particle-hole excitations. The corresponding 
coupling matrix elements are $\sim J^{j+1}/(\hbar\omega)^j$ with $j\ge2$ and the resulting 
avoided crossings are hardly visible in Fig.~\ref{fig:spec}. Generally, the larger $\hbar\omega$
compared to both $U$ and $J$, the more complex are the collective excitations at energies
$\hbar\omega$ and the smaller are the respective coupling matrix elements. 

The formation of an avoided crossing, where the Floquet states of different subspaces $m$ and
$m'$ hybridize, cannot be captured by a perturbation theory describing the system in terms of 
eigenstates labeled by the quantum number $m$. Their presence indicates that the high-frequency 
expansion (\ref{eq:hf}), which can be obtained from such a perturbative approach
\cite{EckardtAnisimovas15}, does not converge. However, we have seen that when both $J$ and $U$ 
are sufficiently small with respect to $\hbar\omega$, the coupling between degenerate states 
originating from different subspaces $m$ will be very small. Therefore, deviations from the
high-frequency approximation will make themselves felt on a large time scale $t_h$ only. 
These deviations can be viewed as heating. In the driven Bose-Hubbard model with strong 
interactions $U\gg J$ discussed in the previous paragraph they correspond to the creation of 
particle-hole excitations \cite{EckardtHolthaus08b,EckardtAnisimovas15}, for weakly interacting 
systems they correspond to energy-non-conserving two-particle scattering
\cite{ChoudhuryMueller14, ChoudhuryMueller15, GenskeRosch15, BilitewskiCooper15, 
BilitewskiCooper15b}. For a numerical study of such heating see \onlinecite{PolettiKollath11}. 

As long as the duration of an experiment is short compared to the heating time $t_h$, it can 
be described by the high-frequency approximation (\ref{eq:hf}). For lattice systems with a 
bound local state space, \emph{e.g.}\ fermionic Hubbard or spin models, it was shown very 
recently that $t_h$ increases exponentially with the driving frequency \cite{AbaninEtAl15, MoriEtAl16, KuwaharaEtAl16} and that the Floquet-Magnus expansion
(\ref{eq:FM}) provides at least an asymptotic expansion for the Floquet Hamiltonian with an 
optimal order $\mu_\text{cut}$ of truncation \cite{KuwaharaEtAl16}. 

The spectrum shown in Fig.~\ref{fig:spec} has been obtained for a small system of five 
particles on five lattice sites only. Approaching the thermodynamic limit, where the system 
size and the particle number are taken to infinity at fixed density, the bands of the spectrum
$\Ho_0$ will approach a continuum and new bands will be created at high energies. The 
quasienergy spectrum of the system, hosting an exponentially large number of quasienergy 
levels in each Brillouin zone (interval of width $\hbar\omega$), will approach a highly 
structured continuum. In this limit, the full effective Hamiltonian will be a very complex 
object, whose eigenstates, the Floquet modes, are superpositions of states having very 
different energies. This scenario explains, why a description in terms of the simple 
expressions provided by the high-frequency approximation~(\ref{eq:hf}) is a suitable approach 
for Floquet engineering, despite the fact that such a description is valid for times $t\ll t_h$
only. 

The existence of a heating time $t_h$ implies that, when a periodically driven system is 
subjected to a quench, \emph{i.e.}\ a sudden change of a parameter like the driving amplitude,
the subsequent relaxation dynamics can consist of two stages. After the quench, the system can
first relax on a time scale $t_r$ to an equilibrium-like state with respect to the effective 
Hamiltonian in high-frequency approximation (\ref{eq:hf}), before the intrinsic heating due to 
the periodic forcing sets in on the time scale $t_h$. Obviously this scenario requires
$t_r\ll t_h$. Such a behavior, which has first been discussed already by \onlinecite{Maricq82}
can be interpreted as a form of \emph{prethermalization}. It is investigated theoretically in 
several recent papers \cite{BukovEtAl15b, CanoviEtAl16, AbaninEtAl15, MoriEtAl16,
KuwaharaEtAl16,AbaninEtAl16b, AbaninEtAl16c}. An interesting effect could be the realization 
of a negative temperature state corresponding to sign-inverted interactions for fermionic 
atoms after a sudden inversion of the effective tunneling matrix element (\ref{eq:Jeff})
\cite{TsujiEtAl11}.

Another intriguing question concerns the relaxation of periodically driven quantum systems on 
time scales much longer than the heating time. From Eq.~(\ref{eq:evolution}) we can infer that 
for a pure state $|\psi(t)\ra$ the time evolution of the expectation value
$O(t)=\la\psi(t)|\hat{O}|\psi(t)\ra$ of an observable $\hat{O}$ can be written like
\be
O(t) = \sum_{n',n} c_{n'}^*c_n e^{\frac{i}{\hbar}(\varepsilon_{n'}-\varepsilon_n)t}
						\la u_{n'}(t)|\hat{O}|u_{n}(t)\ra.
\ee
\onlinecite{RussomannoEtAl12}, have argued that a relaxation to a steady state will correspond 
to the dephasing of the off-diagonal terms with $n'\ne n$, so that asymptotically in the
long-time limit, after a relaxation has occurred, the expectation values are described by 
\be\label{eq:Oasym}
O(t) \simeq \sum_{n} |c_n|^2 \la u_{n}(t)|\hat{O}|u_{n}(t)\ra,
\ee
corresponding to the diagonal ensemble \cite{Dziarmaga10, PolkovnikovEtAl11} with respect to 
the Floquet states. This implies that $O(t)$ becomes time-periodic, $O(t+T)=O(t)$; the system 
synchronizes with the driving. Moreover, \onlinecite{LazaridesEtAl14a} have shown that for
non-interacting (integrable) systems the asymptotic expectation values can be obtained using 
the principle of entropy maximization, under the constraint that the mean occupations
$\la\no_i\ra$ of the single-particle Floquet states $i$ retain their initial values. The 
asymptotic sate is, thus, captured by a periodic generalized Gibbs ensemble, so that for 
typical observables $O(t)\simeq \text{tr}\{\rho(t)\hat{O}\}$ with time-periodic density matrix
$\hat{\rho}(t)\propto\Uo_F(t)\exp(-\sum_i\lambda_i\no_i)\Uo^\dag_F(t)$. The number of 
parameters $\lambda_i$ required to control the integrals of motion $\la\no_i\ra$ grows only 
linearly with the system size. It is much lower than the number of coefficients $c_n$ 
appearing in Eq.~(\ref{eq:Oasym}), which grows exponentially with the system size (since $n$ 
labels the many-body Floquet states, in this case Fock states of the single-particle Floquet 
states, $n=\{n_i\}$). The arguments sketched here in the context of Floquet systems are very 
similar to those employed for the relaxation of isolated autonomous systems \cite{Dziarmaga10, 
PolkovnikovEtAl11}. An interesting exception of the behavior described above are periodically 
driven model systems that in the thermodynamic limit relax to an asymptotic state, dubbed 
\emph{Floquet time crystal}, with $O(t)\ne O(t+T)$, but $O(t)=O(t+kT)$ with integer $k$ 
\cite{ElseEtAl16,vonKeyserlingkSondhi16,vonKeyserlingkEtAl16}. This behavior is associated 
with a spontaneous breaking of the discrete time translational symmetry (similar to the possible
breaking of continuous time translation symmetry proposed recently by \onlinecite{Wilczek12}), 
since the time required to reach a periodic state with $O(t)=O(t+T)$ grows exponentially with 
the system size. 

For non-integrable Floquet systems, it is believed that the system approaches a state described 
by an infinite-temperature ensemble \cite{LazaridesEtAl14b, DAlessioRigol14}. In the sense of 
eigenstate thermalization, for typical observables almost all many-body Floquet modes
$|u_n(t)\ra$ appearing in Eq.~(\ref{eq:Oasym}) are conjectured to give rise to the same
infinite-temperature expectation values for typical observables,
$\la u_n(t)|\hat{O}|u_n(t)\ra\simeq O$ independent of $n$. Roughly speaking, due to the lack 
of energy conservation, the many-body Floquet states are formed by the hybridization of many 
eigenstates of $\Ho_0$ at all available energy scales so that their properties can be computed 
statistically. However, also possible exceptions to this behavior have been discussed
\cite{AbaninEtAl16, LazaridesEtAl15, PonteEtAl15}, including systems featuring many-body 
localization (for which the size of the state space is effectively reduced via the 
segmentation into local subspaces). The work on the relaxation of isolated Floquet systems 
sketched in the last three paragraphs is very recent and it will be interesting to follow 
further developments and possible experimental studies.

\begin{figure}[t]
\includegraphics[width=1\linewidth]{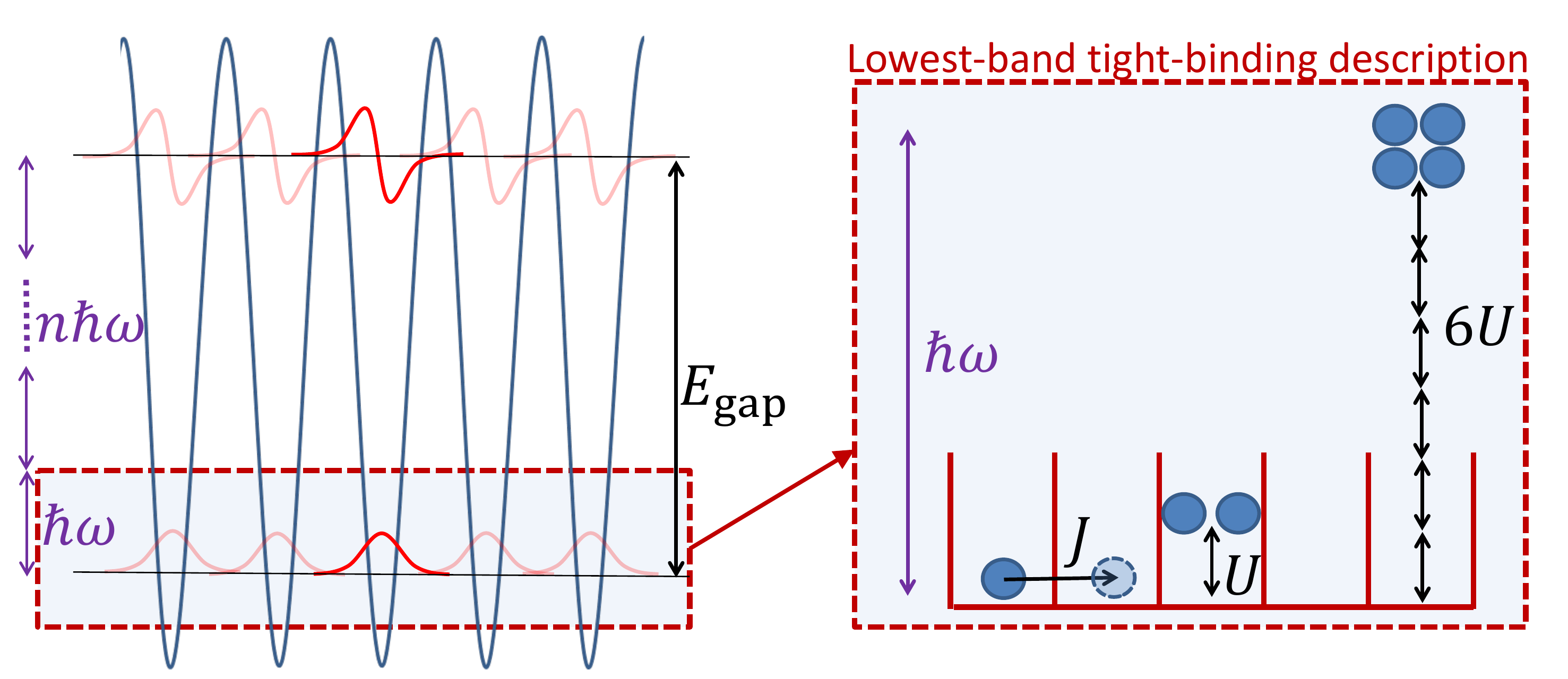}
\centering
\caption{\label{fig:heating} In periodically driven optical lattices, heating occurs due to 
the resonant creation of high-energy excitations, either interband excitations (left) or 
collective intraband-excitations (right, illustrated using the example of a large bosonic site 
occupation). The time scales for these processes should be large compared to the duration of 
the experiment.} 
\end{figure}

Above, we have mentioned that the heating time $t_h$ is expected to increase exponentially with 
the driving frequency for systems with a bound local state space. In experiments with 
ultracold atoms in driven optical lattices, this condition is never fulfilled due to the 
presence of excited orbital states spanning higher-lying Bloch bands (which are not included in 
the low-energy tight-binding description in terms of a Hubbard model). While an effective 
description in terms of low-energy degrees of freedom is very natural in non-driven systems, 
the truncation of high-energy states is a delicate issue in periodically driven systems
already on the single-particle level \cite{HoneEtAl97}. In a periodically driven optical 
lattice the driving frequency is typically chosen such that $\hbar\omega$ lies well below the
band gap $E_\text{gap}$ that separates excited orbital states from the tight-binding state 
space spanned by one low-energy Wannier state in each lattice minimum. However, interband 
excitations can still occur via $n$-``photon'' processes (Fig.~\ref{fig:heating}, left). The 
smaller $n$, the larger will be the coupling matrix element for such interband-heating 
processes. Thus, by increasing the driving frequency, the heating rate associated with the 
resonant creation of collective intraband excitations (\ref{fig:heating}, right), as we 
discussed them above using the example presented in Fig.~\ref{fig:spec}, might decrease. 
However, at the same time the heating rate due to interband excitations tends to increase with 
the driving frequency. Floquet engineering with ultracold atomic quantum gases in optical 
lattice, therefore, requires that there is a window of intermediate frequencies for which 
neither intraband nor interband heating is releavant on the time scale of the experiment. 
Interband transitions can occur both as a consequence of single-particle processes
\cite{DreseHolthaus97, Holthaus15} or two-particle scattering
\cite{ChoudhuryMueller14, ChoudhuryMueller15}. For strong driving, multi-``photon'' interband 
excitations with $n$ as large as nine have recently been observed experimentally and explained 
in terms of single-particle transitions \cite{WeinbergEtAl15}. Arguments based on perturbation 
theory suggest that the rate for such heating processes is suppressed exponentially with
$n\approx E_\text{gap}/(\hbar\omega)$, provided the driving amplitude remains below a 
threshold value \cite{StraeterEckardt16}.

\subsection{\label{sec:EdgeStates}Anomalous topological edge states}
The $\hbar\omega$-periodic structure of the quasienergy spectrum of periodically driven 
quantum systems reflects the interplay between the dynamics occurring within a driving period
(associated with energy scales larger than $\hbar\omega$) with that happening on longer time 
scales (associated with energy scales smaller than $\hbar\omega$). The possibility that 
heating occurs on a long time scale due to the resonant coupling of energetically far distant 
states, discussed in the previous section, is one example of such an interplay. Another, more 
subtle effect related to this interplay is the existence of anomalous topological edge states 
in periodically driven systems \cite{KitagawaEtAl10, JiangEtAl11, KitagawaEtAl12, RudnerEtAl13}.
Without making an attempt to give a complete overview of the numerous recent works on this 
subject, we will briefly sketch the phenomenon in the context of non-interacting spinless 
particles in a two-dimensional lattice, following \onlinecite{RudnerEtAl13}. 

\begin{figure}[t]
\includegraphics[width=1\linewidth]{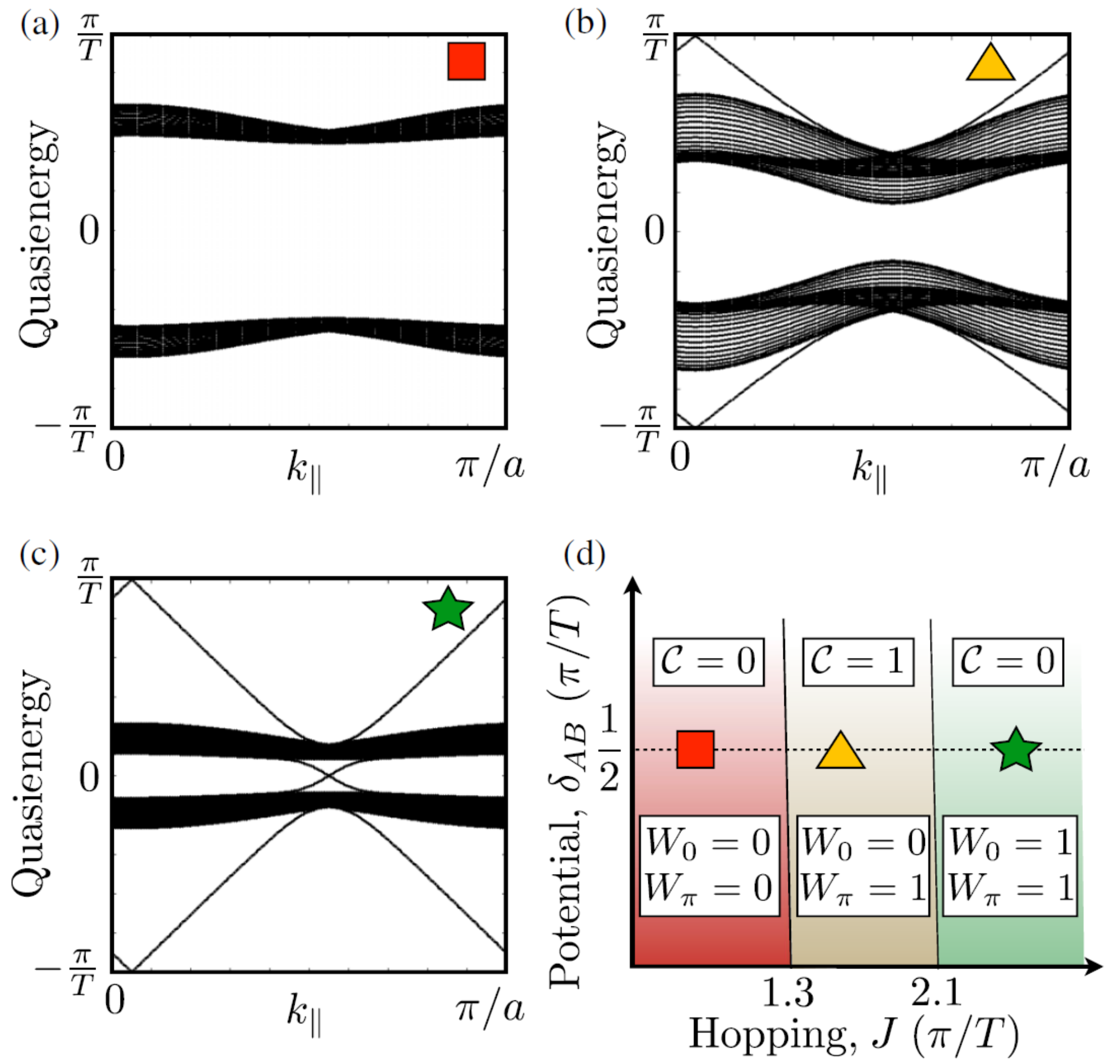}
\centering
\caption{\label{fig:edge} (a-c) Floquet spectra of periodically driven square lattice defined 
in Fig.~\ref{fig:edge2}(a) for the parameters specified in (d). (d) Phase diagram versus 
tunneling parameter $J$ and sublattice off-set $\delta_{AB}$ (in units of $\hbar\omega/2$). 
Phases are characterized by the winding numbers $W_\varepsilon$ for the bulk gaps at 
quasienergy $\varepsilon$; the Chern number of the upper (lower) band is given by $C$ ($-C$).
(taken from \onlinecite{RudnerEtAl13})} 
\end{figure}

Consider a periodically driven two-dimensional tight-binding lattice with $B$ sublattice 
states. For periodic boundary conditions the quasienergy spectrum will possess $B$
Floquet-Bloch bands that shall be separated by gaps. When the translational symmetry is broken 
by open boundary conditions, the system can feature chiral edge states that reflect the 
topological nature of the bulk band structure. Spatially these states are localized at the 
boundary (in the direction perpendicular to it), but delocalized in the direction parallel to 
it. They transport particles in one direction along the boundary only (defined by their
chirality). Figs.~\ref{fig:edge}(a-c) show quasienergy spectra for a driven two-dimensional 
lattice of finite extent with two sublattice states for different parameters. The spectra are 
plotted versus the quasimomentum $k_\parallel$ parallel to two opposite edges
\cite{RudnerEtAl13}. The bulk bands have a finite width according to their dispersion in the 
perpendicular direction. Sometimes neighboring bulk bands are connected by lines traversing a 
band. These lines form one-dimensional bands corresponding to chiral edge states. They come in 
pairs of opposite slope (indicating opposite velocity) corresponding to the two opposite edges. 

Like in autonomous systems, the presence or absence of chiral edge states is connected to the 
topological properties of the bulk (bulk-boundary correspondence, see \emph{e.g}.\ 
\onlinecite{HasanKane10, QiZhang11}). The difference between the number of edge bands entering 
a bulk band $b$ from below and that exiting it above is dictated by the Chern number $C_b$ of 
that band [Eq.~(\ref{eq:chern})], which is a bulk property. However, there is one important 
difference that distinguishes periodically driven from autonomous systems. As a consequence of 
the $\hbar\omega$-periodic structure of the quasienergy spectrum, edge bands can exit the 
uppermost bulk band in the quasienergy interval $[-\hbar\omega/2,\hbar\omega/2]$ above and 
enter the lowermost bulk band from below [Figs.~\ref{fig:edge}(b) and (c)]
\cite{KitagawaEtAl10, JiangEtAl11}. This possibility implies that the system can feature 
chiral edge states even if all bulk bands have Chern number zero, like in
Fig.~\ref{fig:edge}(c). With that it also implies that the presence or absence of chiral edge 
states is not determined by the Chern numbers alone. This is illustrated by the fact that in 
both Figs.~\ref{fig:edge}(a) and (c) the bands have identical Chern numbers, despite the 
respective absence and presence of edge modes. In contrast, in an autonomous systems the 
number of edge states in the gap above a certain bulk band $b$ is given by
$\sum_{\beta\le b} C_\beta$, since no edge bands can enter the lowest band from below.

\onlinecite{RudnerEtAl13}, identify a winding number $W_\varepsilon$ from the bulk properties 
of the system that determines the number of edge modes traversing the band gap containing the 
quasienergy $\varepsilon$, $n_\text{edge}=W_\varepsilon$. The difference
$W_{\varepsilon'}-W_\varepsilon$ corresponds to the sum of the Chern numbers of the bands 
lying between the gaps at $\varepsilon'$ and $\varepsilon$. The winding numbers associated 
with all bulk gaps give a complete topological description of a driven two-dimensional 
lattice. These topological invariants do not only depend on the time-evolution operator over 
one driving period, $\Uo(t_0+T,t_0)$ or $\Uo(T,0)$ for definiteness. They also depend on the 
time-evolution \emph{during} each period as it is captured by the function $\Uo(t)=\Uo(t,0)$, 
that is they depend on the micromotion. The winding number is defined like
\be
W_\varepsilon=\frac{1}{8\pi^2}\int\!\rd t\rd k_x\rd k_y\,
	\text{tr}\Big(\Uo_\varepsilon^\dag\partial_t\Uo_\varepsilon
	\big[\Uo_\varepsilon^\dag\partial_{k_x}\Uo_\varepsilon,
	\Uo_\varepsilon^\dag\partial_{k_y}\Uo_\varepsilon\big]\Big),
\ee
where $\Uo_\varepsilon(\bk,t)$ is a unitary operator with $\Uo_\varepsilon(\bk,T)=1$. It has 
to be obtained by continuously deforming the single-particle time evolution operator
$\Uo(t)=\Uo(\bk,t)$ (in the sector with quasimomentum $\bk$) in a way that the gap at
$\varepsilon$ is smoothly shifted to $\hbar\omega/2$ without being closed on the way. A 
concrete construction of a suitable operator $\Uo_\varepsilon(\bk,t)$ for general
$\Uo(\bk,t)$ is given by \onlinecite{RudnerEtAl13}.

\begin{figure}[t]
\includegraphics[width=1\linewidth]{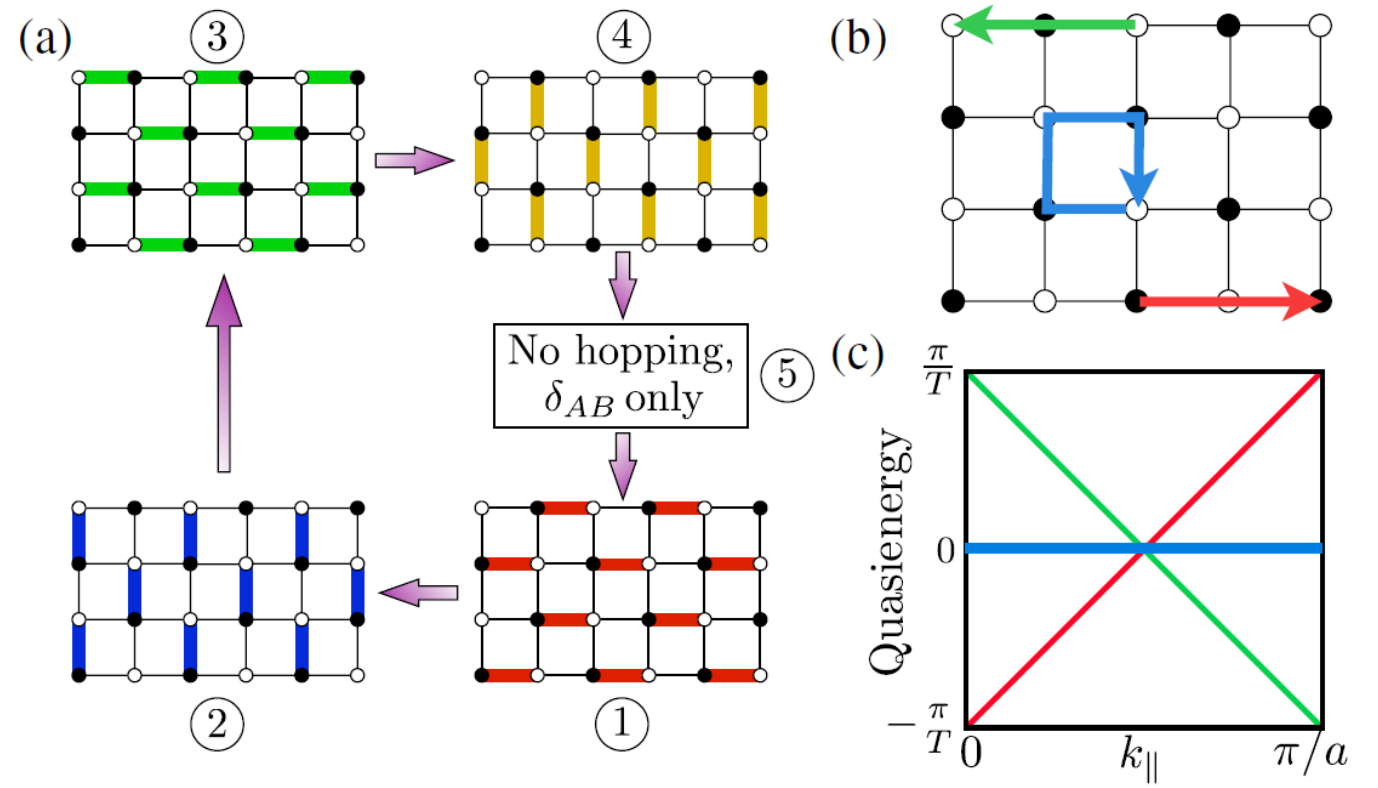}
\centering
\caption{\label{fig:edge2} (a) Simple model system on a square lattice with two sublattice 
states (open and filled circles). Each driving period is divided into five stages of duration
$T/5$. During each stage either tunneling matrix elements $J$ are present on the highlighted 
bonds or an energy offset $\delta_{AB}$ between both sublattices. (b) Dynamics of an initially 
site-localized particle during one driving cycle for fine-tuned parameters
$J=(5/4)\hbar\omega$ and $\delta_{AB}=0$ in the bulk (blue) and at the edges (green, red). (c) 
Quasienergy spectrum for parameters of (b). The bulk levels (blue) are dispersionless, whereas 
the levels for the chiral edge modes (red, green) wrap around the Brillouin zone with constant 
slope. (taken from \onlinecite{RudnerEtAl13}) } 
\end{figure}

The dependence of the winding number on the micromotion has an interesting consequence for the 
bulk-boundary correspondence in Floquet systems: The quasienergy spectrum and the Floquet 
states (including the edge states) can be obtained from the time-evolution operator over one 
driving cycle $\Uo(T)$ by using Eq.~(\ref{eq:Ueigen}). However, the operator $\Uo(T)$ computed 
for a translational invariant system, which represents the bulk properties, does not 
completely determine the properties of the edge states appearing at the boundary of a finite 
system. The edge states depend also on the micromotion of the bulk inherent in the
time-dependence of the evolution operator $\Uo(t)$. 

\onlinecite{RudnerEtAl13}, illustrate this effect using the specific model shown in
Fig.~\ref{fig:edge2}(a). For fine-tuned parameters the dynamics during each driving cycle is 
simply that depicted in Fig.~\ref{fig:edge2}(b): In the bulk, a site-local particle moves 
around a closed circle and returns to its initial state. So while the micromotion in the bulk 
is non-trivial, the Floquet Hamiltonian $\Ho^{F}_0$, which describes/generates the 
stroboscopic dynamics in steps of the driving period $T$, vanishes in the bulk. At the edge of 
the system, the motion around the closed loop is not possible so that a particle is 
transported counter clockwise along the boundary during each driving period, corresponding to 
the formation of a band of chiral edge states. Clearly, this edge dynamics results from the 
interplay of the bulk micromotion with the boundary of the system. The quasienergy spectrum is 
plotted in Fig.~\ref{fig:edge2}(c); while the trivial bulk dynamics over one driving period is 
reflected in a flat bulk dispersion, chiral edge bands with constant slope (representing the 
velocity of two lattice constants per driving period) wrap around the Brillouin zone. 

The model system of Fig.~\ref{fig:edge2}(a) (without driving stage 5, though) and the 
emergence of chiral edge states was investigated recently in experiments with photonic wave 
guides \cite{MukherjeeEtAl16,MaczewskyEtAl16}. An implementation of a similar model defined on 
a hexagonal lattice \cite{KitagawaEtAl11} with ultracold atoms and the observation of chiral 
edge states at interfaces between spatial domains with different topological properties 
has been proposed by \cite{ReichlMueller14}. Anomalous chiral edge states have, moreover, 
been observed in photonic networks \cite{HuEtAl16,GaoEtAl16}. In the circularly forced 
hexagonal lattice \cite{OkaAoki09, KitagawaEtAl11,JotzuEtAl14}, which we discussed in
Sec.~\ref{sec:fti}, anomalous edge states appear for driving frequencies that are low 
enough to resonantly couple the two low-energy Bloch bands described by the Hubbard model
\cite{UsajEtAl14,KunduEtAl14,QuelleEtAl16}, \emph{i.e.}\ when the high-frequency approximation 
(\ref{eq:hf}) breaks down. This is not surprising, since the effective Hamiltonian obtained 
within the high-frequency approximation (\ref{eq:hf}) cannot describe effects related to the
$\hbar\omega$-periodic structure of the quasienergy spectrum, such as the anomalous 
topological edge states discussed in this section.


\subsection{\label{sec:twotime}Two-time formalism}
The concept of Floquet theory can be extended to capture also situations where the Hamiltonian 
is not perfectly periodic in time, e.g.\ when a parameter like the driving strength is varied 
\cite{PfeiferLevine83,BreuerHolthaus89b,PeskinMoiseyev93}. For an arbitrary time-dependent 
Hamiltonian $\Ho(t)$, one can always define a time-periodic Hamiltonian
$\Ho_\tau(t)=\Ho_\tau(t+T)$, with parametric dependence on a second time $\tau$ such that
$\Ho(t)=\Ho_t(t)$. For example, for $\Ho(t)=\Ho_0+p(t)\cos(\omega t)\Vo$, with slowly varying 
amplitude $p(t)$, we can set $\Ho_\tau(t)=\Ho_0+p(\tau)\cos(\omega t)\Vo$. This choice is not 
unique, but for a ``natural'' description of the problem, the dependence on $\tau$ should be 
weak, slow, or limited to a finite interval in time. A suitable description of 
problems where the driving frequency itself varies in time can be obtained by a scaling 
transformation with respect to time \cite{DreseHolthaus99}. 

The quasienergy operator related to $\Ho_\tau(t)$ is given by 
\be
\Qo_\tau(t)=\Ho\tau(t)-i\hbar\rd_t.
\ee
Now a Schr\"odinger-type equation of motion 
\be\label{eq:BH}
i\hbar\rd_\tau |\Psi_\tau\rra = \bar{Q}_\tau|\Psi_\tau\rra
\ee
for states in the extended Floquet Hilbert space can be postulated, where $\bar{Q}_\tau$ 
generates the evolution with respect to the time $\tau$. A straightforward calculation shows 
that 
\be\label{eq:TwoToOne}
|\psi(t)\ra = |\Psi_{\tau}(t)\ra\big|_{\tau=t} \,,
\ee
with $|\Psi_\tau(t)\ra$ being the $\mathcal{H}$-space representation of $|\Psi_\tau\rra$, is a 
solution of the (actual) time-dependent Schr\"odinger equation (\ref{eq:schroedinger}) of the 
Hamiltonian $\Ho(t)$ \cite{BreuerHolthaus89b}. This means that one can integrate
Eq.~(\ref{eq:BH}) in $\mathcal{F}$ in order to compute the time evolution of the system as it 
is described by the time-dependent Schr\"odinger equation (\ref{eq:schroedinger}). The initial 
condition has to obey $|\Psi_{t_0}(t_0)\ra=|\psi(t_0)\ra$, as can be achieved, e.g., by setting
$|\psi_{t_0}\rra = \sum_\alpha \la\psi(t_0)|\alpha\ra \, |\alpha 0 \rra$. The two-time 
formalism provides a Floquet-space description of the time evolution generated by arbitrary 
time-dependent Hamiltonians $\Ho(t)$. Therefore, it constitutes a \emph{Floquet picture}
\cite{BreuerHolthaus89}. Working in the Floquet picture is useful when the Hamiltonian is 
approximately time periodic.

\subsection{\label{sec:adiabatic}Adiabatic state preparation}
A direct consequence of the evolution equation (\ref{eq:BH}) is that one can apply the 
adiabatic principle to Floquet states and their quasienergies \cite{BreuerHolthaus89b}. In 
particular the transition probabilities associated with parameter variations through isolated 
avoided crossings of two quasienergy levels are captured by Landau-Zener theory. While for a 
slow (rapid) parameter variation the crossing is passed adiabatically (diabatically), a 
superposition of both states is created for intermediate rates. 

An important protocol of Floquet engineering is the preparation of the ground state of the 
approximate effective Hamiltonian to be realized by Floquet engineering via a smooth 
parameter variation starting from the ground state of the undriven system. On the level of a 
description in terms of the high-frequency approximation (\ref{eq:hf}), say in leading order
$\Ho_\text{F}\approx\Ho_0$, the ideal dynamics should be adiabatic. However, as was discussed 
in section \ref{sec:heating} using the example of Fig.~\ref{fig:spec}, the ground state of
$\Ho_0$ will undergo avoided level crossings with excited states of energy $m\hbar\omega$. In 
the high-frequency regime these avoided crossings are tiny. For the high-frequency 
approximation to be valid, they should be passed diabatically, reflecting once more that the 
high-frequency approximation is valid on finite times only. Thus, the ideal preparation should 
be based on an \emph{effectively adiabatic} dynamics, defined as a mixture of adiabatic 
dynamics with respect to the high-frequency approximation and diabatic dynamics with respect 
to resonant coupling neglected in this approximation \cite{EckardtHolthaus08b}. Another source 
of heating occurs the effectively adiabatic dynamics guides the system through a phase 
transition. Such a scenario has recently been studied in a spin chain, where
a Kibble-Zurek-like scaling for the creation of excitations has been observed at a
transition induced by resonant coupling \cite{RussomannoDallaTorre16}.

\section{\label{sec:conclusions}Conclusion and outlook}

We have seen that periodic forcing can be powerful tool for the engineering many-body systems 
of ultracold atoms in optical lattices with tailor-made properties. While a basic description 
of such Floquet engineering can often be given in terms of simple (high-frequency single-band)
approximations, the justification of these approximations is a more subtle issue. Future 
research will, therefore, not only be concerned with novel control schemes, but also with the 
stability of Floquet systems towards heating. Efficient strategies for suppressing heating 
will also be crucial for another ambitious goal, the preparation of strongly correlated states 
of matter, such as topologically ordered fractional-quantum-Hall-type states. Also the 
experimental investigation of the relaxation dynamics and the possible prethermalization and 
thermalization of Floquet systems should be a subject of future research. 

Another interesting perspective (going beyond the domain of ultracold quantum gases) is the 
engineering of many-body quantum states of open Floquet systems. When a periodically driven 
quantum system is coupled weakly to a thermal reservoir, it will eventually reach a quasi-
stationary (i.e.\ time periodic) non-equilibrium steady state. The non-equilibrium nature 
results from the fact that the transitions induced by the coupling to the bath do not obey 
detailed balance. Namely a transition $n\to n'$ between two Floquet states with quasienergies 
$\varepsilon_n$ and $\varepsilon_{n'}$ can be accomplished by changing the bath energy by 
$\Delta E_B=\varepsilon_n-\varepsilon_{n'}+m\hbar\omega$, where 
the integer $m$ can take different values. Thus, a particular transition can, for example, 
occur either by lowering or by raising the bath energy. This becomes apparent in
Fermi-golden-rule-type expressions obtained for weak system-bath coupling 
\cite{BluemelEtAl91, KohlerEtAl97, HoneEtAl09, LangemeyerHolthaus14}. The resulting asymptotic
non-equilibrium steady states can have unconventional properties (to mention just a few 
examples, see, e.g.\ recent work by \onlinecite{BreuerEtAl00, TsujiEtAl09, KetzmerickWustmann10,
VorbergEtAl13, VorbergEtAl15, ShiraiEtAl14, FoaTorresEtAl14, SeetharamEtAl15, DehghaniEtAl15, 
GoldsteinEtAl15, IadecolaEtAl15, ShiraiEtAl16}). A powerful tool for the treatment of open 
driven systems is the Floquet-variant of dynamical mean-field theory (see
\onlinecite{AokiEtAl14}, and references therein). Unlike thermal states, non-equilibrium 
steady states are not just determined by a few thermodynamic variables like temperature and 
chemical potential, but depend on the very details of the environment. This makes a 
theoretical treatment challenging, but opens the door for robust dissipative quantum 
engineering of driven many-body systems and their properties. 

\begin{acknowledgments}
During the last decade, the author had the great pleasure to collaborate and discuss with many 
colleagues on the subject of periodically driven quantum systems, including 
Monika Aidelsburger,
Brandon Anderson,
Egidijus Anisimovas,
Ennio Arimondo,
Alejandro Berm\'udez,
Immanuel Bloch,
Alessio Celi,
Donatella Ciampini,
Arnab Das,
Sergey Denisov,
Marco Di Liberto,
Omjyoti Dutta,
Sebastian Greschner,
Peter H\"anggi,
Masudul Haque,
Philipp Hauke,
Andreas Hemmerich,
Gediminas Juzeli\={u}nas,
Roland Ketzmerick,
Sigmund Kohler,
Achilleas Lazarides,
Maciej Lewenstein,
Hans Lignier,
Tania Monteiro,
Cristiane de Morais Smith,
Oliver Morsch,
Takashi Oka,
Christoph \"Olschl\"ager,
Mirta Rodriguez,
Luis Santos,
Ulrich Schneider,
Alexander Schnell,
Klaus Sengstock,
Carlo Sias,
Juliette Simonet,
Ramanjit Sohal,
Fernando Sols,
Parvis Soltan-Panahi,
Tomasz Sowin\'nski,
Shashi Srivastava,
Christoph Str\"ater,
Julian Struck,
Gayong Sun,
Rodolphe Le Targat,
Olivier Tieleman,
Daniel Vorberg,
Malte Weinberg,
Christoph Weiss,
Christof Weitenberg,
Patrick Windpassinger,
Waltraut Wustmann,
Alessandro Zenesini. 
Special thanks go to his teacher Martin Holthaus.

\end{acknowledgments}

\bibliography{mybib}

\end{document}